\documentclass[aps,prd,onecolumn,groupedaddress,showpacs,nofootinbib,amssymb, preprint]{revtex4-2}
\usepackage[dvips]{graphicx}
\usepackage{amssymb}
\usepackage{amsmath}
\usepackage{graphicx,,color}
\usepackage{amsfonts}
\usepackage{bm}
\usepackage{cancel}
\usepackage{comment}
\usepackage{floatflt}
\usepackage{slashed}

\newcommand\e{\mathrm{e}}

\allowdisplaybreaks[4]
\begin{document}

\preprint{KEK-TH-2689, KEK-Cosmo-0372}
\title{Black Holes Thermodynamics \\
and Generalised Non-Extensive Entropy}

\author{Emilio~Elizalde$^1$}\email{elizalde@ice.csic.es}
\author{Shin'ichi~Nojiri$^{2,3}$}\email{nojiri@gravity.phys.nagoya-u.ac.jp}
\author{Sergei~D.~Odintsov$^{1,4}$}\email{odintsov@ice.csic.es}

\affiliation{ $^{1)}$ Institute of Space Sciences (ICE, CSIC) \\ 
C. Can Magrans s/n, 08193 Barcelona, Spain \\
$^{2)}$ Theory Center, High Energy Accelerator Research Organization (KEK), \\
Oho 1-1, Tsukuba, Ibaraki 305-0801, Japan \\
$^{3)}$ Kobayashi-Maskawa Institute for the Origin of Particles and the Universe, \\
Nagoya University, Nagoya 464-8602, Japan \\
$^{4)}$ ICREA, Passeig Lluis Companys, 23, 08010 Barcelona, Spain
}

\begin{abstract}
The first part of this work provides a review of recent research on generalised entropies and their origin, as well as its application to black hole thermodynamics. 
To start, it is shown that the Hawking temperature and the Bekenstein-Hawking entropy are, respectively, 
the only possible thermodynamical temperature and entropy of the Schwarzschild black hole. Moreover, it is investigated if the other known generalised entropies, 
which include R\'enyi's entropy, the Tsallis one, and the four- and five-parameter generalised entropies, could correctly yield the Hawking temperature and the ADM mass. 
The possibility that generalised entropies could describe hairy black hole thermodynamics is also considered, both for the Reissner-Nordstr\"{o}m black hole 
and for Einstein's gravity coupled with two scalar fields. 
Two possibilities are investigated, namely, the case when the ADM mass does not yield the Bekenstein-Hawking entropy, 
and the case in which the effective mass expressing the energy inside the horizon does not yield the Hawking temperature. 
For the model with two scalar fields, the radii of the photon sphere and of the black hole shadow are calculated, which gives constraints on the BH parameters. 
These constraints are seen to be consistent, provided the black hole is of Schwarzschild type. Subsequently, the origin of the generalised entropies is investigated, 
by using their microscopic particle descriptions in the frameworks of a microcanonical and of a canonical ensemble, respectively. 
To finish, the McLaughlin expansion for the generalised entropies is used to derive, in each case, the microscopic interpretation 
of the generalised entropies, via the canonical and the grand canonical ensembles.

\end{abstract}

\maketitle

\newpage

\section{Introduction}\label{SecI}

The thermodynamical properties of gravity could prove to be most important in attempting to construct a theory of quantum gravity. 
Every black hole (BH) can be regarded as a black body with temperature given by the Hawking temperature~\cite{Hawking:1974rv, Hawking:1975vcx} 
and an entropy given by the Bekenstein-Hawking entropy \cite{Bekenstein:1973ur, Hawking:1974rv}. 

In various fields of physics, statistics and informatics, there have been proposed different forms of non-extensive entropies, with their corresponding statistics. 
In particular, the present authors, with some collaborators, have explicitly proposed generalised entropies, which depend on several parameters (see Refs.~\cite{Nojiri:2022aof, Nojiri:2022dkr}). 
They generalise all previously known entropies, as R{\'e}nyi entropy~\cite{renyi}, the Tsallis entropy~\cite{Tsallis:1987eu} (see also \cite{Ren:2020djc, Nojiri:2019skr}), 
the Sharma-Mittal entropy~\cite{SayahianJahromi:2018irq}, Barrow's entropy~\cite{Barrow:2020tzx}, the Kaniadakis entropy~\cite{Kaniadakis:2005zk, Drepanou:2021jiv}, 
Loop Quantum Gravity's entropy~\cite{Majhi:2017zao}, etc. 
Such entropies have been proposed to describe different kinds of physical, statistical, and information systems. 

Note, however, that the Hawking temperature, $T_\mathrm{H}$, can be obtained from the Hawking radiation, which has a thermal distribution. 
This tells us that the Hawking temperature $T_\mathrm{H}$ is independent of the details of the gravity theory, and it is only determined by the geometry. 
Furthermore, if we consider the collapse of the dust shell that yields the black hole, and we assume energy conservation, 
the Arnowitt-Deser-Misner (ADM) mass~\cite{Arnowitt:1959ah} 
must be the thermodynamical energy of the system, at least in the case of a Schwarzschild black hole. 

Recently, a number of works have appeared where different non-extensive kinds of entropies have been applied in the study of black hole thermodynamics (see, e.g, 
\cite{Baruah:2024lyu, Shokri:2024elp, Sekhmani:2024frn, Lu:2024ppa, Xia:2024nmp, Barzi:2024bbj, Anand:2024txo, Nakarachinda:2023jko, 
Gohar:2023hnb, Cimidiker:2023kle, Komatsu:2022bik, Nakarachinda:2022gsb, Manoharan:2022qll, Cimdiker:2022ics, DiGennaro:2022grw, Afshar:2025sav, NooriGashti:2024ywc}). 
Unfortunately, the Hawking temperature or black hole energy obtained in such non-extensive entropy black hole thermodynamics seems to be incorrect. 

One may still conjecture that, in the early universe, the non-extensive generalised entropy could be valid. 
With the universe's evolution, the form of the physical entropy might change to later acquire its current form. 
Therefore, there is some good motivation for the study of different entropies, which were applied in cosmology and BHs. 
In fact, various expressions of entropy lead to different holographic cosmologies~\cite{Nojiri:2021iko, Nojiri:2021jxf} 
and models of holographic dark energy~\cite{Li:2004rb, Li:2011sd, Nojiri:2005pu, Gong:2004cb, Khurshudyan:2016gmb, Landim:2015hqa, Gao:2007ep,Li:2008zq}. 
The holographic approach can be also applied to understand inflation at the early universe~\cite{Nojiri:2019kkp}. 
This makes it possible to describe dark energy and inflation via holographic cosmology in a unified way. 
A microscopic description of the generalised entropy has been also proposed. 
It might be helpful in clarifying the structure of a quantum gravity theory, which is still to be constructed. 

In this review paper, we confirm once more that the Hawking temperature and the ADM mass may correspond to the thermodynamical temperature 
and energy uniquely, at least in the case of the Schwarzschild black hole. 
This shows then that the Bekenstein-Hawking entropy is also a unique BH entropy. 
As a follow-up, we review several approaches where the generalised entropy could be applied for the consideration of several kinds of black holes with hair(s). 

In the next section, we show that the temperature and the entropy of the Schwarzschild black hole are given by the Hawking temperature (Section~\ref{SecIIA}) 
and the Bekenstein-Hawking entropy, respectively, by identifying the ADM mass with the thermodynamical energy (Section~\ref{SecIIB}). 
In Section~\ref{SecIII}, we discuss in more detail the question of whether the Hawking temperature and the Bekenstein-Hawking entropy are unique, or not. 
To this purpose, we show that the ADM mass should be thermodynamical energy, by using the geometry of the black hole and the energy conservation via Birkhoff's theorem. 
In Section~\ref{SecIV}, for the non-extensive entropy, we explicitly consider if it could give the Hawking temperature and the ADM mass correctly. 
In especial, the R\'enyi entropy is discussed in Section~\ref{SecIVA}, Tsallis entropy in Section~\ref{SecIVB}, and further generalised entropies, 
as the four- and five-parameter generalised entropies, in Section~\ref{subge}.
In Section~\ref{SecV}, we study if it might be possible that hairy BH thermodynamics could be described by generalised entropies, for the Reissner-Nordstr\"{o}m black hole 
in Section~\ref{SecVA}, and for Einstein's gravity coupled to two scalar fields, in Section~\ref{SecVB}. 
In the latter case, after showing the general formulation, in Section~\ref{SecVB1}, and some examples, in Section~\ref{SecVB2}, we consider two kinds of possibilities. 
Namely, the case that the ADM mass does not give the Bekenstein-Hawking entropy, in Section~\ref{SecVB3}, and the case 
that the effective mass expressing the energy inside the horizon does not give the Hawking temperature, in Section~\ref{SecVB4}. 
In Section~\ref{generalshadow}, for the models obtained in Section~\ref{SecVB}, we get the radii of the photon sphere and of the black hole shadow. 
Then observations give constraints on the BH parameters. 
They turn out to be consistent, if the black holes are of the Schwarzschild type, although future observations may also give some information about BH thermodynamics. 
In Section~\ref{SecVII}, for more general expressions of the generalised entropies, we propose microscopic particle descriptions of the corresponding thermodynamical system. 
We investigate this problem by using a microcanonical ensemble, in Section~\ref{Sec2}, and a canonical ensemble, in Section~\ref{Sec3}. 
In Section~\ref{SecVIII}, by using the expression of the McLaughlin expansion for the generalised entropies, 
we consider the microscopic interpretation of the generalised entropies in the frame of a canonical ensemble, in Section~\ref{sec-c}, and of a grand canonical ensemble, 
in Section~\ref{sec-gc}. 
The last section of the paper contains a summary and final discussion. 

\section{Entropy Consistent with Hawking Radiation}\label{SecII}

The Hawking radiation has a thermal distribution, from which we can find the Hawking temperature $T_\mathrm{H}$. 
The geometry with the horizon generates Hawking's radiation. 
Therefore, the Hawking temperature, $T_\mathrm{H}$, is only determined by the geometry and is independent of the details of the gravity theory, which realises the geometry. 

Let us consider a system whose size is $R$ and the energy and the entropy inside the system are $E$ and $\mathcal{S}$, respectively. 
Then Bekenstein bound is given by \cite{Bekenstein:1980jp}
\begin{align}
\label{Bbnd}
2\pi R E > \mathcal{S}\, .
\end{align}
In the case of the black hole, $R$ can be identified with the diameter of the horizon, that is, twice the horizon radius. 
We also need to check if the bound (\ref{Bbnd}) is satisfied for general entropy because this bound ensures that the generalised second law of the thermodynamics 
is not violated. 

\subsection{Hawking Temperature from Geometry}\label{SecIIA}

First, we find the Hawking temperature. 
When the metric can be regarded as static, that is, the time-dependence of the metric can be neglected, 
we consider the line element with a horizon at $r=r_\mathrm{H}$, 
\begin{align}
\label{metric}
ds^2 = - P(r) \left( r - r_\mathrm{H} \right) dt^2 + \frac{dr^2}{ P(r) \left( r - r_\mathrm{H} \right) } + r^2 d\Omega_{(2)}^2\, , \quad 
d\Omega_{(2)}^2 \equiv d\vartheta^2 + \sin^2\vartheta d\varphi^2\, \, . 
\end{align} 
Assume that $P(r)$ is positive everywhere and sufficiently smooth in the region near the horizon $r=r_\mathrm{H}$. 
Therefore we may approximate $P(r)$ by a constant, $P(r)\sim P(r_\mathrm{H})$. 
We now introduce a new coordinate $\rho$ defined by 
\begin{align}
d\rho = \frac{dr}{\sqrt{ P(r_\mathrm{H}) \left( r - r_\mathrm{H} \right) }}\, , 
\end{align}
that is, 
\begin{align}
\rho = 2 \sqrt{\frac{r - r_\mathrm{H}}{P(r_\mathrm{H})}}\, . 
\end{align}
By Wick-rotating the time coordinate $t$ as $t\to i\tau$, we obtain the following Euclidean metric 
\begin{align}
ds^2 = \frac{P\left(r_\mathrm{H}\right)^2}{4} \rho^2 d\tau^2 + d\rho^2
+ r(\rho)^2 d\Omega_{(2)}^2 \, . 
\end{align}
We avoid the conical singularity at $\rho=0$ by imposing the periodicity on $\tau$, 
\begin{align}
\frac{P\left(r_\mathrm{H}\right)}{2} \tau \sim \frac{P\left(r_\mathrm{H}\right)}{2} \tau + 2 \pi \, . 
\end{align}
In the finite temperature formalism of the path-integral, the periodicity $\frac{4\pi}{P\left(r_\mathrm{H}\right)}$ corresponds to 
the inverse of the temperature 
\begin{align}
\label{dS6BB}
T_\mathrm{H} = \frac{P\left(r_\mathrm{H}\right)}{4\pi}\, , 
\end{align}
which we call the Hawking temperature. 
In the case of the Schwarzschild spacetime, 
\begin{align}
\label{SchwrzP}
P(r) = P_\mathrm{Schw}(r) \equiv \frac{1}{r}\, , \quad r_\mathrm{H} = 2GM\, .
\end{align}
Here $G$ is Newton's gravitational constant and $M$ is ADM BH mass. 

\subsection{Bekenstein-Hawking Entropy from Thermodynamics}\label{SecIIB}

As is well-known, the area law for the Bekenstein-Hawking entropy \cite{Bekenstein:1973ur} can always be obtained if we identify the 
thermodynamical energy $E$ with the black hole mass $M$, $E=M$, and the temperature of the system with the Hawking temperature (\ref{dS6BB}) 
\cite{Hawking:1975vcx}, 
$T=T_\mathrm{H}=\frac{1}{8\pi GM}$. In fact, the thermodynamical relation $dE= T d\mathcal{S}$ yields
\begin{align}
\label{S1}
d\mathcal{S}=\frac{dE}{T}=8\pi GM dM =d\left( 4 \pi G M^2 \right) \, ,
\end{align}
which can be integrated to be 
\begin{align}
\label{S2}
\mathcal{S}=4 \pi G M^2 + \mathcal{S}_0\,,
\end{align}
where $\mathcal{S}_0$ is a constant of the integration. 
If we assume $\mathcal{S}=0$ when $M=0$, that is, when there is no black hole, we find $\mathcal{S}_0=0$ and we obtain 
\begin{align}
\label{S3}
\mathcal{S}= \frac{\pi {r_\mathrm{H}}^2}{G} = \frac{A}{4G} \, .
\end{align}
Here $A\equiv 4\pi {r_\mathrm{H}}^2$ is the area of the horizon. 
Therefore the Bekenstein-Hawking entropy, that is, the area law for BH entropy, can be obtained by assuming $E=M$ and 
$T=T_\mathrm{H}$ by using the thermodynamical relation $d\mathcal{S}=dE/T$.
Note that the Bekenstein-Hawking entropy $\mathcal{S}$, of course, satisfies the Bekenstein bound in (\ref{Bbnd}) 
because $2\pi R E = 4 \pi r_\mathrm{H} \frac{r_\mathrm{H}}{2G} = 2 \mathcal{S}>\mathcal{S}$. 

\section{Uniqueness of Hawking Temperature and Bekenstein-Hawking Entropy}\label{SecIII}

We now consider whether the Hawking temperature and the Bekenstein-Hawking entropy could be unique or not. 
For this purpose, we need to consider the following two points, 
\begin{enumerate}
\item Can the thermodynamical energy $E$ be identified with the black hole mass $M$ ({\em i.e.}, $E=M$)? 
\item Is the temperature of the black hole given by the Hawking temperature, $T=T_\mathrm{H}$?
\end{enumerate}
For the first point, we should be careful in the following situation, that is if BH is not the Schwarzschild one nor isolated one, 
there is no Arnowitt-Deser-Misner mass. 
Then the mass $M$ may be the quasilocal mass contained in the horizon sphere or given by the ``black hole part'' of the spacetime. 
For several quasilocal mass prescriptions, see Ref.~\cite{Szabados:2009eka} for a review.

\subsection*{$E=M$?}\label{EMgeneral}\label{SecIIIA}

To consider the first point, the following `thought experiments' could be useful. 
\begin{enumerate}
\item We assume an infalling spherically symmetric shell of dust with mass $M$ and the initial radius sufficiently large. 
The Birkhoff theorem~\cite{Wald:1984rg} tells that the spacetime outside the shell is the Schwarzschild one~(\ref{SchwrzP}). 
The mass $M$ is nothing but the mass of the shell. 
Inside the shell, the spacetime is empty and flat. 
\item By the collapse of the shell, the radius becomes smaller and smaller. 
A black hole is formed when the shell crosses the Schwarzschild radius $r_\mathrm{H}=2M$ in (\ref{SchwrzP}). 
\item The resulting geometry is always asymptotically flat and the shell mass $M$ appearing in the horizon radius is surely the energy $E$ of the system, 
$E=M$ because the energy should be conserved during the collapse of the shell due to the Birkhoff theorem. 
That is, the geometry outside of the shell does not change during the collapse. 
Therefore the energy of the final black hole must be the mass of the shell. 
We should note that due to spherical symmetry, the quadrupole does not appear and the gravitational waves, which might carry the energy, 
are not emitted during the collapse. 
\end{enumerate}

One may consider other definitions of the mass or the energy of the black hole like 
the Misner-Sharp-Hernandez quasilocal mass $M_\mathrm{MSH}$ defined in any 
spherically symmetric spacetime by \cite{Misner:1964je,Hernandez:1966zia} 
and the Brown-York quasilocal energy \cite{Brown:1992br}. 
The obtained results are consistent with the above arguments or totally unphysical (for more detailed arguments, see \cite{Nojiri:2021czz}). 

\subsection{$T=T_\mathrm{H}$?}\label{SecIIIB}

The second point is discussed in the previous section. 
As mentioned there, Hawking radiation is obtained if the geometry with the horizon is prescribed 
and the standard Hawking temperature is the parameter appearing as the temperature in the thermal distribution of the emitted Hawking radiation. 
We may imagine that we put the black hole in a heat bath at temperature $T$. 
Then the thermal equilibrium between the black hole radiation and the heat bath occurs when the radiation temperature 
equals the temperature of the heat bath, $T=T_\mathrm{H}$. 
Therefore, the heat bath can be used as a thermometer and the temperature measured by the 
heat bath must be the standard Hawking temperature of the Hawking radiation and, therefore, we find $T=T_\mathrm{H}$. 

\section{Consistency of General Entropies}\label{SecIV}

Due to some motivations, different kinds of entropy other than the Bekenstein-Hawking one \cite{Bekenstein:1973ur, Hawking:1975vcx} have been proposed 
like Tsallis~\cite{Tsallis:1987eu}, R\'{e}nyi~\cite{renyi}, Barrow~\cite{Barrow:2020tzx}, Sharma-Mittal~\cite{SayahianJahromi:2018irq}, Kaniadakis~\cite{Kaniadakis:2005zk} 
and loop quantum gravity entropies~\cite{Majhi:2017zao}). 
Furthermore, generalised entropy with three, four, five and six parameters has been proposed in~\cite{Nojiri:2022aof, Nojiri:2022dkr, Odintsov:2022qnn
}. 
These generalised entropies give all the aforementioned known entropies within a certain choice of entropic parameters. 

\subsection{R\'enyi Entropy}\label{SecIVA}

First, we consider the R\'enyi entropy 
\cite{Czinner:2015eyk, Tannukij:2020njz, Promsiri:2020jga, Samart:2020klx} 
\begin{align}
\label{RS1}
\mathcal{S}_\mathrm{R}=\frac{1}{\alpha} \ln \left( 1 + \alpha \mathcal{S} \right) \, .
\end{align} 
Here $\mathcal{S}$ is the Bekenstein-Hawking entropy~(\ref{S3}) and $\alpha$ is a parameter specifying the deformation from the Bekenstein-Hawking entropy. 
In the limit of $\alpha\rightarrow 0$, the expression (\ref{RS1}) reduces to the Bekenstein-Hawking entropy. 
By using Eq.~(\ref{S2}) with $\mathcal{S}_0=0$, we find, 
\begin{align}
\label{RS2}
\mathcal{S}_\mathrm{R}=\frac{1}{\alpha} \ln \left( 1 + 4 \pi \alpha G M^2 \right) \, .
\end{align}
Note the R\'enyi entropy satisfies the Bekenstein bound (\ref{Bbnd}) because $\mathcal{S}_\mathrm{R}<\mathcal{S}<2\pi R E$ 
as long as $\mathcal{S}_\mathrm{R}>0$. 

\subsubsection{Assumption $M=E$}\label{SecIVA1}

If the mass $M$ coincides with the energy $E$ of the system due to the energy conservation 
\cite{Czinner:2015eyk,Tannukij:2020njz,Promsiri:2020jga,Samart:2020klx}, 
the consistency of the system with the thermodynamical equation $d\mathcal{S}=dE/T$ requires to define 
the ``R\'enyi temperature'' $T_\mathrm{R}$ by 
\begin{align}
\label{TR1}
\frac{1}{T_\mathrm{R}} \equiv \frac{d\mathcal{S}_\mathrm{R}}{dM} 
= \frac{8 \pi G M}{1 + 4 \pi \alpha G M^2} \,,
\end{align}
that is, 
\begin{align}
\label{TR2}
T_\mathrm{R} = \frac{1}{8\pi GM} + \frac{\alpha M}{2} = T_\mathrm{H} 
+ \frac{\alpha}{16 \pi G T_\mathrm{H}} \, ,
\end{align}
which is different from the Hawking temperature $T_\mathrm{H}$ and therefore the ``R\'enyi temperature'' $T_\mathrm{R}$ is not 
the temperature perceived by any observer detecting Hawking radiation, as we stressed. 
Hence the ``R\'enyi temperature'' $T_\mathrm{R}$ could be physically irrelevant for black hole thermodynamics. 

\subsubsection{$T=T_\mathrm{H}$?}\label{SecIVA2}

Instead of assuming that the thermodynamical energy $E$ is identical with BH mass $M$, 
we now assume that the thermodynamical temperature $T$ coincides with the Hawking temperature $T_\mathrm{H}$. 

By using the thermodynamical relation $dE=Td\mathcal{S}$, the assumptions $T=T_\mathrm{H}$ and $\mathcal{S}=\mathcal{S}_\mathrm{R}$ 
show that the corresponding thermodynamical energy $E_\mathrm{R}$ is given by 
\begin{align}
dE_\mathrm{R} =&\, T_\mathrm{H} d\mathcal{S}_\mathrm{R} = \frac{1}{8\pi GM} \frac{8 \pi GM dM}{1 + 4\pi \alpha G M^2} 
= \frac{dM}{1 + 4\pi \alpha G M^2} \, , 
\label{RE1}
\end{align}
which can be integrated to give, 
\begin{align}
\label{RE2}
E_\mathrm{R} = \frac{\arctan \left( \sqrt{4\pi \alpha G} M \right) }{\sqrt{4\pi \alpha G}} 
= M - \frac{4\pi \alpha GM^2 }{3} + \mathcal{O}\left( \alpha^2 \right) \, .
\end{align}
Here the integration constant is fixed so that $E_\mathrm{R}=0$ when $M=0$. 
The correction $- \frac{4\pi \alpha GM^2 }{3} + \mathcal{O}\left( \alpha^2 \right)$ shows that the expression~(\ref{RE2}) of the 
thermodynamical energy $E_\mathrm{R}$ is different from BH mass $M$, $E_\mathrm{R}\neq E$, 
what looks unphysical. 
The more important thing is that it seems to conflict with energy conservation when we consider the spherically symmetric 
dust shell collapses to a Schwarzschild black hole. 

\subsection{Tsallis Entropy}\label{SecIVB}

Let us consider Tsallis entropy \cite{Tsallis:1987eu} in BH thermodynamics as is discussed in \cite{Nojiri:2021czz}. 

The Tsallis entropy may be considered as an alternative 
to the Bekenstein-Hawking entropy \cite{Ren:2020djc} (see also \cite{Nojiri:2019skr}), 
\begin{align}
\label{TS1}
\mathcal{S}_\mathrm{T} = \frac{A_0}{4G} \left( \frac{A}{A_0} \right)^\delta \, .
\end{align}
Here $A_0$ is a constant with the dimension of the area and $\delta$ specifying the non-extensivity. 
In the limit of $\delta \to 1$, the expression in (\ref{TS1}) reduces to the standard Bekenstein-Hawking entropy~(\ref{S3}). 
Note, however, that the Bekenstein bound (\ref{Bbnd}) is violated for the large black hole because 
$\frac{\mathcal{S}_\mathrm{T}}{\mathcal{S}} \to \infty$ when $\mathcal{S} \to \infty$ if $\delta>1$. 

\subsubsection{$M=E$?}\label{SecIVB1}

Again by assuming that the thermodynamical energy $E$ is given by BH mass $M$, 
we obtain $A=4\pi \left( 2GM \right)^2 = 16\pi G^2 E^2$ and the expression in (\ref{TS1}) has the following form, 
\begin{align}
\label{TS2}
\mathcal{S}_\mathrm{T} = \frac{A_0}{4G} \left( \frac{16\pi G^2 E^2}{A_0} 
\right)^\delta \, , 
\end{align}
which may allow us to define ``Tsallis temperature'' as follows, 
\begin{align}
T_\mathrm{T} \equiv \frac{dE}{d\mathcal{S}_\mathrm{T}}
= \frac{2G}{\delta A_0 E^{2\delta - 1}} \left( \frac{A_0}{16\pi G^2} \right)^\delta 
= \frac{2G}{\delta A_0 M^{2\delta - 1}} \left( \frac{A_0}{16\pi G^2} \right)^\delta \, .
\label{TS3}
\end{align}
The Tsallis temperature is, of course, different from the Hawking temperature (\ref{dS6BB}), 
$T=T_\mathrm{H}=\frac{1}{8\pi GM}$ unless $\delta =1$. 

\subsubsection{$T=T_\mathrm{H}$?}\label{SecIVB2}

Instead of identifying the black hole mass $M$ with the thermodynamical energy $E$, 
we now assume that the BH temperature is the Hawking temperature. 
Because we have $A= 4\pi \left( 4\pi T_\mathrm{H} \right)^{-2} = \frac{1}{4\pi {T_\mathrm{H}}^2}$, we find
\begin{align}
\label{TS4}
\mathcal{S}_\mathrm{T} = \frac{{A_0}^{1-\delta}}{4 G \left(4\pi {T_\mathrm{H}}^2 \right)^\delta} \, ,
\end{align}
which may lead to the ``Tsallis energy'' $E_\mathrm{T}$ given by 
\begin{align}
\label{TS5}
d E_\mathrm{T} = T_\mathrm{H} d\mathcal{S}_\mathrm{T} 
= - \frac{\delta {A_0}^{1-\delta} d T_\mathrm{H}}{2 G \left(4\pi\right)^\delta {T_\mathrm{H}}^{2\delta}} \, .
\end{align}
By integrating (\ref{TS5}), we obtain 
\begin{align}
\label{TS6}
E_\mathrm{T} = \frac{\delta {A_0}^{1-\delta} }{2 \left( 2\delta -1 \right) G \left(4\pi\right)^\delta {T_\mathrm{H}}^{2\delta-1}} 
= \frac{\delta {A_0}^{1-\delta} \left( 8\pi GM \right)^{2\delta -1} }{2 \left( 2\delta -1 \right) G \left(4\pi\right)^\delta } \, .
\end{align}
Here we have fixed the integration constant by imposing the condition that $E_\mathrm{T}=0 $ when $M=0$. 
The standard relation $E_\mathrm{T} = M$ is reproduced for $\delta=1$ when the Tsallis entropy reduces to the Bekenstein-Hawking entropy. 

\subsubsection{Bekenstein-Hawking Entropy as Tsallis Entropy}\label{SecIVB3}

The standard thermodynamics is related to the extensive system. 
In the system, if we separate the system with thermodynamical energy $E$ into two systems with $E_1$ and $E_2$ with $E=E_1 + E_2$, 
the standard entropy $\mathcal{S}_\mathrm{standard}(E)$ is extensive, that is, 
\begin{align}
\label{standardentropy}
\mathcal{S}_\mathrm{standard} \left(E_1 + E_2\right) 
= \mathcal{S}_\mathrm{standard} \left(E_1\right) + \mathcal{S}_\mathrm{standard} \left(E_2\right) \, .
\end{align}
On the other hand, the original Tsallis entropy $\tilde{\mathcal{S}}_\mathrm{T}$ has the following properties, 
\begin{align}
\label{Tsalllis}
\left( \tilde{\mathcal{S}}_\mathrm{T} \left(E_1 + E_2\right) \right)^\frac{1}{\delta} 
= \left( \tilde{\mathcal{S}}_\mathrm{T} \left(E_1\right) \right)^\frac{1}{\delta} + \left( \tilde{\mathcal{S}}_\mathrm{T} \left(E_2\right) \right)^\frac{1}{\delta} \, .
\end{align}
As pointed in \cite{Volovik:2024wif}, the standard Bekenstein-Hawking entropy is recovered 
 with $\delta=2$
\begin{align}
\label{BH}
\left( \mathcal{S} \left(E_1 + E_2\right) \right)^\frac{1}{2} 
= \left( \mathcal{S} \left(E_1\right) \right)^\frac{1}{2} 
+ \left( \mathcal{S}_\mathrm{T} \left(E_2\right) \right)^\frac{1}{2} \, .
\end{align}
As claimed in \cite{Volovik:2024wif}, this property could be explained by the quantum process where a black hole 
could split into smaller black holes. 

We should note that the black hole is not in equilibrium with the heat bath or environment. 
It is like indoor bright red charcoal. The black hole is hotter than the vacuum. 
The non-extensivity of the Bekenstein-Hawking entropy in (\ref{BH}) could tell that the internal energy could not be extensive, either. 
In the Tsallis entropy, long-range force is supposed to generate non-extensivity because the long-range force makes the internal energy non-extensive. 

\subsection{Generalised Entropies}\label{subge}

The generalised four- and six-parameter generalised entropies have the following forms \cite{Nojiri:2022aof, Nojiri:2022dkr},
\begin{align}
\mathcal{S}_4 \left(\alpha_{\pm},\delta,\gamma\right) = \frac{1}{\gamma}\left[\left(1 + \frac{\alpha_+}{\delta} \mathcal{S}\right)^{\delta}
 - \left(1 + \frac{\alpha_-}{\delta} \mathcal{S}\right)^{-\delta}\right] \, ,
\label{intro-1}
\end{align}
and
\begin{align}
\mathcal{S}_6 \left(\alpha_{\pm},\delta_{\pm},\gamma_{\pm}\right) = \frac{1}{\alpha_+ + \alpha_-}
\left[ \left( 1 + \frac{\alpha_+}{\delta_+} \mathcal{S}^{\gamma_+} \right)^{\delta_+} 
 - \left( 1 + \frac{\alpha_-}{\delta_-} \mathcal{S}^{\gamma_-} \right)^{-\delta_-} \right]\, ,
\label{intro-2}
\end{align}
respectively. 
Here $\mathcal{S} = \frac{A}{4G}$ represents the Bekenstein-Hawking entropy~(\ref{S3}). 
Both of these entropies reduce to all the aforementioned known entropies for a suitable limit of the respective parameters, that is, 
Tsallis, R\'{e}nyi, Barrow, Sharma-Mittal, Kaniadakis, and loop quantum gravity entropies. 
For instance, we find
\begin{itemize}
\item $\mathcal{S}_4$ reduces to the Tsallis entropy in the limit of $\alpha_{+} \rightarrow \infty$, $\alpha_{-} = 0$ and $\gamma = \left(\alpha_{+}/\beta\right)^{\beta}$. 
\item the six parameter entropy, $\mathcal{S}_6$ goes to the Tsallis entropy for $\alpha_+ = \alpha_- \rightarrow 0$ and $\gamma_+ = \gamma_-$. 
\end{itemize}
In addition to the four- and six-parameter generalised entropies, a three-parameter entropy was also proposed in \cite{Nojiri:2022aof} in the following form
\begin{align}
\mathcal{S}_3\left(\alpha,\delta,\gamma\right) = \frac{1}{\gamma}\left[\left(1 + \frac{\alpha}{\delta} \mathcal{S}\right)^\delta - 1\right] \, .
\label{intro-3}
\end{align}
$\mathcal{S}_3$ cannot be, however, reduced to the Kaniadakis entropy in any parameter limit. 
Therefore the four-parameter entropy is the minimal generalisation because the minimum number of parameters required 
in an entropy function for generalising all the known entropies is four. 
In $\mathcal{S}_3$, $\mathcal{S}_4$, and $\mathcal{S}_6$, the Bekenstein bound (\ref{Bbnd}) can be violated in some parameter regions because 
they have limits where these entropies go to the Tsallis entropy, where the Bekenstein bound is violated for the large black hole if $\delta>1$. 

We should also note that $\mathcal{S}_3$, $\mathcal{S}_4$, and $\mathcal{S}_6$ share the following properties: 
\begin{enumerate}
\item They obey the third law of thermodynamics, i.e., they vanish in the limit of $\mathcal{S} \rightarrow 0$. 
\item They are monotonically increasing functions of the variable $\mathcal{S}$. 
\item They diverge in the limit $\mathcal{S} \to \infty$. 
\end{enumerate}
For the last point, when we consider the cosmology, $A$ is given by the area of the apparent horizon, $A=\frac{4\pi}{H^2}$. 
Here $H$ is the Hubble rate. 
Therefore $\mathcal{S}_3$, $\mathcal{S}_4$, and $\mathcal{S}_6$ diverge when the Hubble rate vanishes, $H = 0$ 
because the Bekenstein-Hawking entropy $\mathcal{S}$ itself diverges at $H = 0$. 
This singular behaviour is common to all the known entropies like the Tsallis, the R\'{e}nyi, the Barrow, the Kaniadakis, the Sharma-Mittal and the loop quantum gravity entropy. 

In order to solve the problem of the singularity when $H\to 0$, a five-parameter entropy was proposed in \cite{Odintsov:2022qnn}, 
which has the following form, 
\begin{align}
\mathcal{S}_5\left(\alpha_{\pm},\delta,\gamma,\epsilon\right) 
= \frac{1}{\gamma}\left[\left\{1 + \frac{1}{\epsilon}\tanh\left(\frac{\epsilon \alpha_+}{\delta} \mathcal{S}\right)\right\}^{\delta}
 - \left\{1 + \frac{1}{\epsilon}\tanh\left(\frac{\epsilon \alpha_-}{\delta} \mathcal{S}\right)\right\}^{-\delta} \right] \, .
\label{intro-4}
\end{align}
Due to $\tanh$ function, the entropy (\ref{intro-4}) does not show the singularity even if $\mathcal{S}$ diverges or $H\to 0$. 
This entropy, therefore, admits a bouncing scenario, where $H$ vanishes at the bouncing time. 

In the following, for the generalised entropies $\mathcal{S}_4$ in (\ref{intro-1}) and $\mathcal{S}_5$ in (\ref{intro-4}), we investigate 
if the mass $M$ coincides with the thermal energy $E$ by assuming that the temperature $T$ is given by the Hawking temperature $T_\mathrm{H}$, $T=T_\mathrm{H}$, 
and also inversely, if the temperature $T$ is given by the Hawking temperature $T_\mathrm{H}$ by assuming the mass $M$ coincides with the thermal energy $E$, $E=M$. 

\subsubsection{$E=M$?}\label{SecIVC1}

\noindent
{\bf The case of four-parameter generalised entropy $\mathcal{S}_4$}

By substituting $\mathcal{S}=4\pi GM^2$ in (\ref{S2}) with $\mathcal{S}_0=0$ to the four-parameter generalised entropy in (\ref{intro-1}), we find 
\begin{align}
dE_4 = T_\mathrm{H} d \mathcal{S}_4
= \frac{1}{\gamma}\left[\left(1 + \frac{4\pi \alpha_+}{\delta} GM^2\right)^{\delta-1} \alpha_+ 
+ \left(1 + \frac{4\pi \alpha_-}{\delta} GM^2\right)^{-\delta-1} \alpha_- \right] dM \, .
\label{intro-1dE}
\end{align}
Here $E_4$ is the energy defined by the first relation $dE_4 = T_\mathcal{H} d \mathcal{S}_4$. 
The above expression does not give $dE_4=dM$ nor $E_4=M$ in general, of course. 

When $M$ is small, Eq.~(\ref{intro-1dE}) gives, 
\begin{align}
\label{dE4expnd}
dE_4 \sim \frac{\alpha_+ + \alpha_-}{\gamma}dM \, ,
\end{align}
which can be integrated to be 
\begin{align}
\label{dE4expndint}
E_4 \sim E_{4(0)} + \frac{\alpha_+ + \alpha_-}{\gamma}M \, ,
\end{align}
Here $E_{4(0)}$ is a constant of the integration. 
Eq.~(\ref{dE4expndint}) tells $E_4 \neq M$ in general but if we choose 
\begin{align}
\label{dE4expndintc}
\frac{\alpha_+ + \alpha_-}{\gamma} = 1\, , \quad E_{4(0)} =0\, ,
\end{align}
we obtain $E_4 = M$. 

On the other hand, when $M$ is large, if we choose $\delta>0$, we obtain 
\begin{align}
\label{dE4l}
E_4 \sim E_{4(1)} + \frac{\alpha_+ M}{\gamma \left( 2\delta - 1 \right) } \left(\frac{4\pi \alpha_+}{\delta} GM^2\right)^{\delta -1} \, .
\end{align}
Here $E_{4(1)}$ is a constant of the integration. 
Anyway, Eq.~(\ref{dE4l}) generally gives $E_4 \neq M$ but 
if we choose 
\begin{align}
\label{dE4le}
\delta=1\, , \quad \frac{\alpha_+}{\gamma} =1 \, , \quad E_{4(1)} = 0\, ,
\end{align}
we obtain $E_4 = M$. 

Note that the condition (\ref{dE4le}) is compatible with the condition (\ref{dE4expndint}) if 
\begin{align}
\label{cmpt}
\alpha_+ = \gamma \, , \quad \alpha_- = 0\, , \quad \delta=1 \, ,
\end{align}
and we obtain an expression of the entropy which realises $E_4 = M$ in both of the limits $M\to 0$ and $M\to +\infty$. 
The condition (\ref{cmpt}), however, shows that the four-parameter generalised entropy $\mathcal{S}_4$ in (\ref{intro-1}) reduces 
to the standard Bekenstein-Hawking entropy, $\mathcal{S}_4\to \mathcal{S}$.

\ 

\noindent{\bf The case of five-parameter generalised entropy $\mathcal{S}_5$}

In the case of the five-parameter generalised entropy in (\ref{intro-1}), we find 
\begin{align}
dE_5 =&\, T_\mathrm{H} d \mathcal{S}_5\left(\alpha_{\pm},\delta,\gamma,\epsilon\right) \nonumber \\
=&\, \frac{1}{\gamma}\left[\left\{1 + \frac{1}{\epsilon}\tanh\left(\frac{4\pi \epsilon \alpha_+}{\delta} GM^2 \right)\right\}^{\delta-1} 
\frac{\alpha_+}{\cosh^2 \left(\frac{4\pi \epsilon \alpha_+}{\delta} GM^2 \right)} \right. \nonumber \\
& \left. \quad + \left\{1 + \frac{1}{\epsilon}\tanh \left(\frac{4\pi \epsilon \alpha_-}{\delta} GM^2 \right)\right\}^{-\delta-1} 
\frac{\alpha_- }{\cosh^2 \left(\frac{4\pi \epsilon \alpha_-}{\delta} GM^2 \right)}
\right] dM \, .
\label{intro-4dE}
\end{align}
Here $E_5$ is the energy defined by $dE_5 = T_\mathcal{H} d \mathcal{S}_5$ and 
the above expression tells $dE_5\neq dM$ nor $E_5 \neq M$ in general. 

When $M$ is small, one again obtains (\ref{dE4expnd}) and (\ref{dE4expndint}). 
The obtained result tells $E_4 \neq M$ again in general but if we choose the parameters as in (\ref{dE4expndintc}), 
we obtain $E_4 = M$. 

When $M$ is large, by assuming $\alpha_+ > \alpha_->0$, we find 
\begin{align}
\label{5lM}
dE_5 \sim \frac{4\alpha_-}{\gamma} \left( 1 + \frac{1}{\epsilon} \right)^{-\delta-1} \exp \left( - \frac{8\pi \epsilon \alpha_-}{\delta} GM^2 \right) dM\, .
\end{align}
The integration of the above equation is given by using Gauss' error function $\mathrm{erf}$, which is defined by 
\begin{align}
\label{erf}
\mathrm{erf}(x) \equiv \frac{2}{\sqrt{\pi}} \int_0^x \e^{-t^2} dt \, ,
\end{align} 
as follows
\begin{align}
\label{5lMint}
E_5 \sim E_{5(0)} + \frac{1}{\gamma} \left( 1 + \frac{1}{\epsilon} \right)^{-\delta-1} 
\sqrt{\frac{\alpha_- \delta}{2 \epsilon G}} \mathrm{erf} \left( M \sqrt{\frac{8\pi \epsilon \alpha_- G}{\delta}} \right) \, .
\end{align}
Therefore there is no choice of the parameter which reproduces $E_4 = M$ except the limit that 
$\mathcal{S}_5$ in (\ref{intro-4}) reduces to the standard Bekenstein-Hawking entropy, $\mathcal{S}_5\to \mathcal{S}$.

\subsubsection{$T=T_\mathrm{H}$?}\label{SecIVC2}

In the case of the four-parameter generalised entropy in (\ref{intro-1}), if we identify the thermodynamical energy $E$ with the mass $M$, 
the corresponding temperature $T_4$ is given by 
\begin{align}
\label{fourT}
T_4 \equiv \frac{d \mathcal{S}_4}{dM}
= \frac{8\pi GM}{\gamma}\left[\left(1 + \frac{4\pi \alpha_+}{\delta} GM^2\right)^{\delta-1} \alpha_+ 
+ \left(1 + \frac{4\pi \alpha_-}{\delta} GM^2\right)^{-\delta-1} \alpha_- \right] \, .
\end{align}
Eq.~(\ref{fourT}) does not give the Hawking temperature $T_\mathrm{H}$, $T_4\neq T_\mathrm{H}=\frac{1}{8\pi GM}$ 
besides the limit that $\mathcal{S}_4$ in (\ref{intro-1}) reduces to the standard Bekenstein-Hawking entropy, $\mathcal{S}_4\to \mathcal{S}$. 

On the other hand, in the case of the five-parameter generalised entropy in (\ref{intro-4}), we obtain 
\begin{align}
\label{fiveT}
T_5 \equiv \frac{d \mathcal{S}_5}{dM} 
= \frac{8\pi GM}{\gamma} &\, \left[\left\{1 + \frac{1}{\epsilon}\tanh\left(\frac{4\pi \epsilon \alpha_+}{\delta} GM^2 \right)\right\}^{\delta-1} 
\frac{\alpha_+}{\cosh^2 \left(\frac{4\pi \epsilon \alpha_+}{\delta} GM^2 \right)} \right. \nonumber \\
& \left. + \left\{1 + \frac{1}{\epsilon}\tanh \left(\frac{4\pi \epsilon \alpha_-}{\delta} GM^2 \right)\right\}^{-\delta-1} 
\frac{\alpha_- }{\cosh^2 \left(\frac{4\pi \epsilon \alpha_-}{\delta} GM^2 \right)}
\right] \, .
\end{align}
Eq.~(\ref{fiveT}) does not give the Hawking temperature $T_\mathrm{H}$, $T_5\neq T_\mathrm{H}=\frac{1}{8\pi GM}$, either. 

\section{More General Black Hole}\label{SecV}

The thermodynamical relation $dE=Td\mathcal{S}$ does not generally hold, for example, if there is a chemical potential. 
The first law of thermodynamics is, 
\begin{align}
\label{T1}
dE = dQ + dW\, .
\end{align}
$dQ=Td\mathcal{S}$ is the heat which flows into the system and $dW$ is the work which the system received. 
The variation of the work $dW$ can be expressed as 
\begin{align}
\label{T1B}
dW=-P dV + \sum_i \mu_i dN_i\, .
\end{align}
Here $P$ and $V$ are the pressure and the volume of the system and $dN_i$ is the number of the $i$-th kind of particles which flow into the system and 
$\mu_i$ is the corresponding chemical potential. 

When we discussed if the thermodynamical energy should be the ADM mass in Section~\ref{EMgeneral} by using the falling dust shell, 
we have assumed that the region outside the dust shell is the vacuum. 
In a realistic situation, all the matter does not fall into the black hole but the matter outside the horizon contributes to the ADM mass. 
In the case of the Reissner-Nordstr\"{o}m black hole, the ADM mass includes the contributions from the electromagnetic field outside the horizon. 
More in general, if BH has any hair, the energy density of the hair contributes to the ADM mass and changes the thermodynamical relation 
$dE=Td\mathcal{S}$ as in (\ref{T1}). 
In this section, we discuss the possibility that the generalised entropies could be given by the hairy black hole. 
We now review the thermodynamics of the Reissner-Nordstr\"{o}m black hole, and after that, we consider the black hole with scalar hair(s). 
For the construction of the black hole with scalar hair(s), we use the model where the Einstein gravity couples with two scalar fields. 

\subsection{Reissner-Nordstr\"{o}m Black Hole}\label{SecVA}

The metric of the Reissner-Nordstr\"{o}m BH is given by the following line element, 
\begin{align}
\label{RNmetric}
ds^2 = - \left( 1 - \frac{2GM}{r} + \frac{GQ^2}{r^2} \right) dt^2 + \frac{dr^2}{1 - \frac{2GM}{r} + \frac{GQ^2}{r^2}} + r^2 d\Omega_{(2)}^2 \, . 
\end{align}
Here $Q$ is the electric charge of the black hole and the ADM mass is given by $M$ as in the Schwarzschild black hole. 
As well-known, the Reissner-Nordstr\"{o}m black hole has two horizons. 
The radii $r_\pm$ of the horizons are given by 
\begin{align}
\label{RNhorizon}
r_\pm = GM \pm \sqrt{G^2 M^2 - G Q^2}\, , 
\end{align}
Here $r_+$ is the radius of the outer horizon and $r_-$ is that of the inner one. 
Eq.~(\ref{RNhorizon} shows that the Bekenstein-Hawking entropy $\mathcal{S}$ is given by, 
\begin{align}
\label{RNent}
\mathcal{S}_\pm = \frac{\pi {r_\pm}^2}{G} 
= \frac{\pi \left( G M \pm \sqrt{G^2 M^2 - G Q^2} \right)^2}{G} \, . 
\end{align}
Here $\mathcal{S}_+$ is the entropy corresponding to the outer horizon and $\mathcal{S}_-$ to the inner one. 
The Hawking temperature $T_\mathrm{H}$ corresponding to the outer horizon is given by 
\begin{align}
\label{RNTH}
T_\mathrm{H} = \frac{\sqrt{G^2 M^2 - G Q^2} }{2\pi \left( G M + \sqrt{G^2 M^2 - G Q^2} \right)^2} \, ,
\end{align}
Then we find 
\begin{align}
\label{RNMQ}
T_\mathrm{H} d\mathcal{S}_+ 
= dM - \frac{Q}{GM + \sqrt{G^2 M^2 - G Q^2}} dQ
\end{align}
Then there is a correction by the last term. 

One may consider a possibility to define a generalised entropy $\mathcal{S}_\mathrm{g}$ instead of (\ref{RNMQ}), 
\begin{align}
\label{T1gRN}
T_\mathrm{H} d \mathcal{S}_\mathrm{g} = dM \, .
\end{align}
In the case of the Reissner-Nordstr\"{o}m black hole, it is generally impossible because the system depends on two variables $M$ and $Q$. 
Let first assume $\mathcal{S}_\mathrm{g}$, $\mathcal{S}_\mathrm{g}=\mathcal{S}_\mathrm{g} (M,Q)$. 
Then Eq.~(\ref{T1gRN}) can be rewritten as, 
\begin{align}
\label{T1gRN2}
T_\mathrm{H} \left( \frac{\partial \mathcal{S}_\mathrm{g}}{\partial M}dM + \frac{\partial \mathcal{S}_\mathrm{g}}{\partial Q}dQ \right) = dM \, .
\end{align}
Then we find $\frac{\partial \mathcal{S}_\mathrm{g}}{\partial Q}=0$ and therefore the integrablity condition requires $\frac{\partial T_\mathrm{H}}{\partial Q}$ 
because $\frac{\partial \mathcal{S}_\mathrm{g}}{\partial M}$ should not depend on $Q$. 
This conflicts with the expression of the Hawking temperature in (\ref{RNTH}), which explicitly depends on $Q$. 
A possibility is to consider a one-dimensional line in the two-dimensional $M$-$Q$ as $Q=Q(M)$. 
Then Eq.~(\ref{T1gRN2}) tells, 
\begin{align}
\label{SgRN}
\mathcal{S}_\mathrm{g} (M) \int \frac{dM}{T_\mathrm{H} \left( M, Q\left(M\right) \right)}\, .
\end{align}
As an example, we consider the case $Q=q_0 M$ with a constant satisfying a condition ${q_0}^2 < G$. 
In this case, Eq.~(\ref{RNTH}) gives 
\begin{align}
\label{RNTHg}
T_\mathrm{H} = \frac{\sqrt{G^2 - G {q_0}^2} }{2\pi \left( G + \sqrt{G^2 - G {q_0}^2} \right)^2M} \, ,
\end{align}
and therefore Eq.~(\ref{SgRN}) can be integrated to give, 
\begin{align}
\label{SgRNg}
\mathcal{S}_\mathrm{g} = \frac{\pi \left( G + \sqrt{G^2 - G {q_0}^2} \right)^2M^2}{\sqrt{G^2 - G {q_0}^2}} \, .
\end{align}
Here we choose the constant of the integration so that $\mathcal{S}_\mathrm{g}$ vanishes when $M$ vanishes. 
The obtained expression (\ref{SgRNg}) is proportional to $M^2$, which is similar to the Bekenstein-Hawking entropy 
in (\ref{S2}) with $\mathcal{S}_0=0$ although the coefficient is different. 
Other function $Q=Q(M)$ gives more general expressions but it depends on the physical process of the black hole creation. 
The case $Q=q_0 M$ could correspond to the process that BH is created only by the accretion of the charged particle 
whose ratio of the charge with the mass is $q_0$. 

\subsection{Gravity Coupled with Two Scalar Fields}\label{SecVB}

In \cite{Nojiri:2020blr}, it has been shown that arbitrarily given spherically symmetric spacetimes can be realised 
within Einstein's gravity coupled with two scalar fields even if the spacetime is time-dependent. 
The original model of Ref.~\cite{Nojiri:2020blr}, however, includes ghosts, which make the model inconsistent. 
After that, it was found that the ghosts could be excluded by imposing constraints by the Lagrange multiplier 
fields~\cite{Nojiri:2023dvf}. 

The action in the model of Ref.~\cite{Nojiri:2020blr} includes two scalar fields $\phi$ and $\chi$, which couple with Einstein's gravity, 
\begin{align}
\label{I8}
S_{\mathrm{GR} \phi\chi} = \int d^4 x \sqrt{-g} &\, \left[ \frac{R}{2\kappa^2}
 - \frac{1}{2} A (\phi,\chi) \partial_\mu \phi \partial^\mu \phi
 - B (\phi,\chi) \partial_\mu \phi \partial^\mu \chi \right. \nonumber \\
& \left. \qquad - \frac{1}{2} C (\phi,\chi) \partial_\mu \chi \partial^\mu \chi
 - V (\phi,\chi) + \mathcal{L}_\mathrm{matter} \right]\, .
\end{align}
Here $A(\phi,\chi)$, $B(\phi,\chi)$, and $C(\phi,\chi)$ are called kinetic functions and $V(\phi,\chi)$ is the potential, which are functions 
of the two scalar fields $\phi$ and $\chi$. 
Furthermore, $\mathcal{L}_\mathrm{matter}$ is the Lagrangian density of matter. 
The gravitational coupling constant $\kappa$ is related to Newton's gravitational constant $G$ as $\kappa^2=8\pi G$. 
In this section, we mainly use the geometrised units $c=G=1$. 

General spherically symmetric and time-dependent spacetime is described by the metric given by the following line element,
\begin{align}
\label{GBiv_time}
ds^2 = - \e^{2\nu (t,r)} dt^2 + \e^{2\lambda (t,r)} dr^2 + r^2 d\Omega_{(2)}^2 \, .
\end{align}
We also assume, 
\begin{align}
\label{TSBH1}
\phi=t\, , \quad \chi=r\, ,
\end{align}
which does not lead to any loss of generality~~\cite{Nojiri:2020blr}. 

We should note, however, that the functions $A$ and/or $C$ are often negative, which makes $\phi$ and/or $\chi$ to be ghosts. 
The ghosts can be eliminated by imposing constraints by using the Lagrange multiplier fields $\lambda_\phi$ and $\lambda_\chi$ 
and modifying the action (\ref{I8}) $S_{\mathrm{GR} \phi\chi} \to S_{\mathrm{GR} \phi\chi} + S_\lambda$, 
where the additional term $S_\lambda$ is given by
\begin{align}
\label{lambda1}
S_\lambda = \int d^4 x \sqrt{-g} \left[ \lambda_\phi \left( \e^{-2\nu(t=\phi, r=\chi)} \partial_\mu \phi \partial^\mu \phi + 1 \right)
+ \lambda_\chi \left( \e^{-2\lambda(t=\phi, r=\chi)} \partial_\mu \chi \partial^\mu \chi - 1 \right) \right] \, .
\end{align}
By varying $S_\lambda$ with respect to $\lambda_\phi$ and $\lambda_\chi$, we obtain the following constraints:
\begin{align}
\label{lambda2}
0 = \e^{-2\nu(t=\phi, r=\chi)} \partial_\mu \phi \partial^\mu \phi + 1 \, , \quad
0 = \e^{-2\lambda(t=\phi, r=\chi)} \partial_\mu \chi \partial^\mu \chi - 1 \, ,
\end{align}
which is consistent with the assumption (\ref{TSBH1}). 
The constraints from Eq.~(\ref{lambda2}) make the scalar fields $\phi$ and $\chi$ non-dynamical, and the fluctuations of
$\phi$ and $\chi$ around the background (\ref{TSBH1}) do not propagate ( 
see \cite{Nojiri:2023dvf}
for detail). 

We now construct a model which has a solution realising the functions $\e^{2\nu(t,r)}$ and $\e^{2\lambda(t,r)}$ in Eq.~(\ref{GBiv_time}). 
The matter is assumed to be a perfect fluid with the energy density $\rho$ and the pressure $p$, 
\begin{align}
\label{FRk2}
T_{\mathrm{matter}\, tt} =-g_{tt}\rho\ ,\quad T_{\mathrm{matter}\, ij}=p\, g_{ij}\, .
\end{align}
Here $i, j= r,\vartheta, \varphi$. 
For the spacetime given by Eq.~(\ref{GBiv_time}), the Einstein equations can re rewritten as follows, \begin{align}
\label{ABCV}
A=& \frac{\e^{2\nu}}{\kappa^2} \left\{ - \e^{-2 \nu} \left[ \ddot\lambda + \left( \dot\lambda - \dot\nu \right) \dot\lambda \right]
+ \e^{-2\lambda}\left[ \frac{\nu' + \lambda'}{r} + \nu'' + \left( \nu' - \lambda' \right) \nu' + \frac{\e^{2\lambda} - 1}{r^2}\right] \right\} \nonumber \\
&\, - \e^{2\nu} \left( \rho + p \right) \, , \nonumber \\
B=&\, \frac{2\dot\lambda}{\kappa^2 r} \, , \nonumber \\
C=&\, \frac{\e^{2\lambda}}{\kappa^2} \left\{ \e^{-2 \nu} \left[ \ddot\lambda + \left( \dot\lambda - \dot\nu \right) \dot\lambda \right]
 - \e^{-2\lambda}\left[ - \frac{\nu' + \lambda'}{r} + \nu'' + \left( \nu' - \lambda' \right) \nu' + \frac{\e^{2\lambda} - 1}{r^2}\right] \right\} 
\, , \nonumber \\
V=& \frac{\e^{-2\lambda}}{\kappa^2} \left( \frac{\lambda' - \nu'}{r} + \frac{\e^{2\lambda} - 1}{r^2} \right) - \frac{1}{2} \left( \rho - p \right) \, .
\end{align}
This tells that we obtain a model that realises the spacetime described by the metric (\ref{GBiv_time}) by finding $(t,r)$-dependence of $\rho$ and $p$ and 
by replacing $(t,r)$ in Eq.~(\ref{ABCV}) with $(\phi,\chi)$.

\subsubsection{Black Hole with Scalar Hair}\label{SecVB1}

We now consider the time-independent geometry, that is, static, spherical, and asymptotically flat spacetimes, 
\begin{align}
\label{metricB}
ds^2 = g_{\mu\nu} dx^{\mu} dx^{\nu} 
= -\e^{2\nu(r)} dt^2 + \e^{2\lambda(r)} dr^2 + r^2 d\Omega_{(2)}^2 \, .
\end{align}
Asymptotic flatness corresponds to $\lim_{r\to +\infty} 
\lambda(r)=0$ and we normalise the time coordinate $t$, to $\lim_{r\to+\infty} \nu( r )=0$. 

Let us now investigate the effects of the scalar hair and write the energy density of the scalar fields by $\rho$. 
Then as in the standard Tolman-Oppenheimer-Volkov (TOV) equation, the time-time component of the Einstein equations gives
\begin{align}
\label{EinTOV0}
 - \kappa^2 \rho = \frac{1}{r^2} \left( r \e^{-2\lambda} - r \right)' \, .
\end{align}
Here a prime ``$'$'' denotes differentiation with respect to $r$. 
The mass function $m(r)$ is defined by 
\begin{align}
\label{massfn}
\e^{-2\lambda} \equiv 1 - \frac{2G m(r)}{r} \,,
\end{align}
which gives $4\pi r^2 \rho = m'(r) $ and by integrating the expression, we obtain 
\begin{align}
\label{TOV1}
m(r) =4 \pi \int^r r^{\prime 2} \rho (r') dr' \,. 
\end{align}
In the case of a compact star like a neutron star, the lower limit of the integration is chosen to be $r=0$. 
In the case of the black hole, the boundary condition is given at the horizon $r=r_\mathrm{H}$ so that 
\begin{align}
\label{boundaryhorizon}
2G m\left(r_\mathrm{H} \right) = r_\mathrm{H}\, .
\end{align}
If the geometry is asymptotically Schwarzschild spacetime, the ADM mass is given by 
\begin{align}
\label{mass}
M=m(r\to\infty)=4 \pi \int^\infty d r \, r^2 \rho(r) \,.
\end{align}
Note that $m(r=\infty)$ is not the total mass, which should be defined by 
\begin{align}
\bar{M} =&\, \int d^3 x \, \sqrt{\gamma} \, \rho(r) 
= 4 \pi \int_0^\infty \rho(r) r^2 \e^{\lambda(r)} d r 
= 4 \pi \int_0^\infty \rho(r) r^2 \left[1-\frac{2 G m(r)} r\right]^{-1/2} d r \nonumber\\
=&\, 4 \pi \int_0^{\infty} dr \, \rho(r) r^2 \left[1 + \frac{G m(r)} r - \frac{3G^2 m^2(r)}{r^2} + \mathcal{O} \left( G^3 \right) \right] \, .
\label{barM}
\end{align} 
Here $\gamma$ is the determinant of the three-dimensional spatial metric, 
\begin{align}
\label{induced3metric}
\gamma_{\ell m} \, d x^{\ell} d x^m =\e^{2 \lambda} d r^{2}+r^{2} 
d\Omega_{(2)}^2 \, .
\end{align}
The second term in the last line of Eq.~(\ref{barM}) can be interpreted as the Newtonian gravitational potential energy 
\begin{align}
\label{NP}
 - 4 \pi G \int_0^\infty dr \, \rho(r) \, r^2 \frac{m(r)} r = - \frac{G}{2} 
\int dV \int dV' \, \frac{\rho\left(\bm r\right) \rho\left(\bm r'\right)}{\left| \bm r - \bm r' \right|} \, .
\end{align}
Here $dV$ and $dV'$ are three-dimensional volume elements and the 
general-relativistic nonlinear corrections are identified by 
$G^2$ term and higher power terms of $G$.

The above arguments could tell that the contribution to the mass from the scalar hair could be given by 
\begin{align}
\label{hairmass}
M_\mathrm{hair} = m\left( r=\infty \right) - m\left( r=r_\mathrm{H} \right)\, .
\end{align}
This term gives a correction as in the second term of Eq.~(\ref{T1})
\begin{align}
\label{correction}
dM=T_H d\mathcal{S} + dM_\mathrm{hair}\, .
\end{align}
Then the correction of the general entropy from the Bekenstein-Hawking entropy might be interpreted as the contribution from $M_\mathrm{hair}$. 
As we can identify $M=m\left( r=\infty \right)$, however, Eq.~(\ref{correction}) can be rewrittten as 
\begin{align}
\label{correctionB}
dm\left( r=r_\mathrm{H} \right)=T_H d\mathcal{S} \, .
\end{align}
Because $m\left( r=r_\mathrm{H} \right)=\frac{r_\mathrm{H}}{2G}$ and the Hawking temerature is given by 
$T_\mathrm{H}=\frac{1}{4\pi r_\mathrm{H}}$, Eq.~(\ref{correctionB}) is approved only if we choose $\mathcal{S}$ to be the Bekenstein-Hawking entropy, 
$\mathcal{S}=\frac{\pi {r_\mathrm{H}}^2}{G}$ as in the standard black hole thermodynamics. 

We should note, however, that there might be a possibility to define a generalised entropy $\mathcal{S}_\mathrm{g}$ as in (\ref{T1gRN})
by using the first law in (\ref{T1}) as follows
\begin{align}
\label{T1g}
T_\mathrm{H} d \mathcal{S}_\mathrm{g} \equiv dQ + dW = dE \, .
\end{align}
We investigate the possibility in the following. 

Even for more general gravity theories including the modified gravities, as an analogue of (\ref{EinTOV0}), 
we may define the effective energy density $\rho_\mathrm{eff}$ by using only geometry, 
\begin{align}
\label{MGTOV}
 - \kappa^2 \rho_\mathrm{eff} = \frac{1}{r^2} \left( r \e^{-2\lambda} - r \right)' \,,
\end{align}
and also define the effective mass function as in (\ref{massfn})
\begin{align}
\label{massfneff}
\e^{-2\lambda} \equiv 1 - \frac{2G m_\mathrm{eff}(r)}{r} \, .
\end{align}
Then by integrating (\ref{MGTOV}), we obtain the counterpart of (\ref{TOV1}), 
\begin{align}
\label{MDmassfunc}
m_\mathrm{eff}(r)=4 \pi \int^r_0dr' r^{\prime 2} \rho_\mathrm{eff} (r')\, .
\end{align}
We may interpret $m_\mathrm{eff}(r)$ as the mass acted upon by the attractive force at radius $r$. 
We use this definition later. 

\subsubsection{Examples}\label{SecVB2}

In order to consider the examples, we now assume \cite{Nojiri:2022sfd}, 
\begin{align}
\label{exam1}
\e^{2\nu}= \e^{-2\lambda}= \frac{1}{h_2(r)} \left( 1 - \frac{r_\mathrm{H}}{r} \right) \, , 
\end{align}
with a constant radius of the horizon $r_\mathrm{H}$. 
We do not include matter besides the two scalar fields $\phi$ and $\chi$. 
Then the expressions (\ref{ABCV}) give, 
\begin{align}
\label{ABCVexphi}
A(\phi)=&\, \frac{1}{\kappa^2 h_2(\phi)} \left( 1 - \frac{r_\mathrm{H}}{\phi} \right) 
\left\{ - \frac{h_2(\phi)h_2''(\phi) - 2 {h_2'(\phi)}^2}{2{h_2(\phi)}^3} \left( 1 - \frac{r_\mathrm{H}}{\phi} \right) - \frac{r_\mathrm{H}h_2'(\phi)}{\phi^2{h_2(\phi)}^2} \right. \nonumber \\
&\, \left. + \frac{1}{\phi^2} \left( 1 - \frac{1}{h_2(\phi)} \right) \right\} \, , \nonumber \\
B(\phi)=&\, 0 \, , \nonumber \\
C(\phi)=&\, - \frac{ h_2(\phi)}{\kappa^2\left( 1 - \frac{r_\mathrm{H}}{\phi} \right)}
\left\{ - \frac{h_2(\phi) h_2''(\phi) - 2 {h_2'(\phi)}^2}{2{h_2(\phi)}^3} \left( 1 - \frac{r_\mathrm{H}}{\phi} \right) - \frac{r_\mathrm{H}h_2'(\phi)}{\phi^2{h_2(\phi)}^2} \right. \nonumber \\
&\, \left. + \frac{1}{\phi^2} \left( 1 - \frac{1}{h_2(\phi)} \right) \right\} \, , \nonumber \\
V(\phi)=& \frac{1}{\kappa^2} \left\{ \frac{h_2'(\phi)}{\phi {h_2(\phi)}^2} \left( 1 - \frac{r_\mathrm{H}}{\phi} \right) 
+ \frac{1}{\phi^2} \left( 1 - \frac{1}{h_2(\phi)} \right) \right\} \, .
\end{align}
We should note that $A$, $C$, and $V$ in (\ref{ABCVexphi}) depend explicitly on the horizon radius $r_\mathrm{H}$, 
that is, the horizon radius is fixed in this model. 
There could be other solutions besides Eq.~(\ref{exam1}), but it could not be easy to find them. 
This problem can be bypassed by using the trick of Ref.~\cite{Nojiri:2017kex}. 
We add a new term in the Lagrangian density including new fields $\sigma$ and $\rho_\mu$ 
as $\mathcal{L}_{\rho\sigma}=\rho^\mu \partial_\mu \sigma$. 
By the variation of $\mathcal{L}_{\rho\sigma}$ with respect to $\rho^\mu$ 
yields constant $\sigma$,
\begin{align}
\label{rhosigma2}
\partial_\mu \sigma =0 \,,
\end{align}
We now identify $\sigma $ with the horizon radius $r_\mathrm{H} $. 
By replacing $r_\mathrm{H}$ with $\sigma $ in the equations in (\ref{ABCVexphi}), 
$r_\mathrm{H}$ is given as an integration constant appearing from Eq.~(\ref{rhosigma2}), 
\begin{align}
\label{ABCVexphisigma}
A(\phi,\sigma)=&\, \frac{1}{\kappa^2 h_2(\phi,\sigma)} \left( 1 - \frac{\sigma }{\phi} \right) 
\left\{ - \frac{h_2(\phi,\sigma)h_{2,\phi\phi}(\phi,\sigma) - 2 {h_{2,\phi}(\phi,\sigma)}^2}{2{h_2(\phi,\sigma)}^3} 
\left( 1 - \frac{\sigma }{\phi} \right) \right. \nonumber \\
&\, \left. \qquad \qquad \qquad \qquad 
 - \frac{\sigma h_{2,\phi}(\phi,\sigma)}{\phi^2{h_2(\phi,\sigma)}^2} 
+ \frac{1}{\phi^2} \left( 1 - \frac{1}{h_2(\phi,\sigma)} \right) \right\} \, , \nonumber \\
B(\phi,\sigma)=&\, 0 \, , \nonumber \\
C(\phi,\sigma)=&\, - \frac{ h_2(\phi,\sigma)}{\kappa^2\left( 1 - \frac{\sigma }{\phi} \right)}
\left\{ - \frac{h_2(\phi,\sigma) h_{2,\phi\phi}(\phi,\sigma) - 2 {h_{2,\phi}(\phi,\sigma)}^2}{2{h_2(\phi,\sigma)}^3} 
\left( 1 - \frac{\sigma }{\phi} \right) \right. \nonumber \\
&\, \left. \qquad \qquad \qquad \qquad 
 - \frac{\sigma h_{2,\phi}(\phi,\sigma)}{\phi^2{h_2(\phi,\sigma)}^2} 
+ \frac{1}{\phi^2} \left( 1 - \frac{1}{h_2(\phi,\sigma)} \right) \right\} \, , \nonumber \\
V(\phi,\sigma)=& \frac{1}{\kappa^2} \left\{ \frac{h_{2,\phi}(\phi,\sigma)}{\phi {h_2(\phi,\sigma)}^2} \left( 1 - \frac{\sigma }{\phi} \right) 
+ \frac{1}{\phi^2} \left( 1 - \frac{1}{h_2(\phi,\sigma)} \right) \right\} \, .
\end{align}
Here $h_2(\phi,\sigma)_{,\phi} \equiv \partial h_2(\phi,\sigma) / \partial\phi $, 
$h_2(\phi,\sigma)_{,\phi\phi} \equiv \partial^2 h_2(\phi,\sigma) / \partial\phi^2$. 
By the choice of $h_2$, we obtain several examples. 

\subsubsection{Thermodynamics}\label{SecVB3}

As an example, we consider the case
\begin{align}
\label{h2}
h_2 = 1 + \frac{2G M \left( r_\mathrm{H} \right) - r_\mathrm{H}}{r} = 1 + \frac{2G M(\sigma) - \sigma}{\phi} \, .
\end{align}
Then when $r$ is large, Eq.~(\ref{exam1}) tells
\begin{align}
\label{exam1B}
\e^{2\nu}= \e^{-2\lambda} \sim 1 - \frac{2G M\left(r_\mathrm{H}\right)}{r} \, .
\end{align}
Therefore $M\left( r_\mathrm{H} \right)$ is the ADM mass. 

In order to consider the possibility of (\ref{T1g}), as an example, we consider the R\'enyi entropy in (\ref{RS1}), which has 
now the following form 
\begin{align}
\label{RS1ex}
\mathcal{S}_\mathrm{R}=\frac{1}{\alpha} \ln \left( 1 + \alpha \frac{\pi {r_\mathrm{H}}^2}{G} \right) \, .
\end{align} 
Because the Hawking temperature is given by 
$T_\mathrm{H}=\frac{1}{4\pi r_\mathrm{H}}$, if we assume (\ref{T1g}), 
we find 
\begin{align}
\label{RS1ex2}
M'\left( r_\mathrm{H} \right) d r_\mathrm{H} 
= \frac{1}{2G \left( 1 + \alpha \frac{\pi {r_\mathrm{H}}^2}{G} \right)} d r_\mathrm{H} 
= \frac{1}{2\sqrt{\pi \alpha G}}d \left( \mathrm{Arctan} \left( r_\mathrm{H} \sqrt{\frac{\pi \alpha}{G}} \right) \right) \, .
\end{align}
Therefore in (\ref{h2}) if we choose
\begin{align}
\label{RSex3}
M(\sigma) = \left. \frac{1}{2\sqrt{\pi \alpha G}}d \mathrm{Arctan} \left( r_\mathrm{H} \sqrt{\frac{\pi \alpha}{G}} \right) \right|_{r_\mathrm{H}=\sigma}\, ,
\end{align}
we obtain a model whose entropy is described by the R\'enyi entropy $\mathcal{S}_\mathrm{R}$. 

Similarly, for the generalised entropy $\mathcal{S}_\mathrm{g}=\mathcal{S}_\mathrm{g} \left( r_\mathrm{H} \right)$, if we choose $M(\sigma)$ in (\ref{h2}) by 
\begin{align}
\label{gnrlM}
M(\sigma) = \int^\sigma dr_\mathrm{H} \frac{\mathcal{S}_\mathrm{g}' \left( r_\mathrm{H} \right)}{4\pi r_\mathrm{H}} \, ,
\end{align}
a model whose entropy is $\mathcal{S}_\mathrm{g}$ can be constructed. 

\subsubsection{Thermodynamics Based on $m_\mathrm{eff}$}\label{SecVB4}

Here based on \cite{Nojiri:2022sfd}, we consider the thermodynamics by using $m_\mathrm{eff}$ in (\ref{MDmassfunc}). 
Instead of (\ref{exam1}), we assume, 
\begin{align}
\label{h3}
\e^{-2\lambda(r)}=\e^{2\nu(r)} h_3(r)\, .
\end{align}
Here $h_3(r)$ is a positive function of $r$. 
As in (\ref{ABCVexphisigma}), the geometry (\ref{h3}) is realised by using (\ref{ABCV}) 
with the Lagrangian density $\mathcal{L}_{\rho\sigma}=\rho^\mu \partial_\mu \sigma$, 
\begin{align}
\label{TSBH6sCABCD} 
A(\phi,\sigma)=&\,\frac{1}{\kappa^2{h_3(\phi,\sigma)^2}} \left( \frac{\sigma}{\phi} -1 \right)^2 
\left[ - \frac{ h_{3,\phi}(\phi,\sigma) }{4\phi h_3(\phi,\sigma)} + \frac{3 h_{3,\phi}(\phi,\sigma) }
{4\left(\phi - \sigma\right) h_3(\phi,\sigma)} \right. \nonumber\\
& \left. \qquad \qquad \qquad \qquad \qquad 
+ \frac{h_{3,\phi\phi}(\phi,\sigma) }{2 \, h_3(\phi,\sigma)} - \frac{1}{4}\left(\frac{h_{3,\phi}(\phi,\sigma) }{h_3(\phi,\sigma)}\right)^2 \right] \, , \nonumber \\
B(\phi,\sigma)=&\, 0 \, , \nonumber \\
C(\phi,\sigma)=&\, \frac{1}{\kappa^2} \left[ \frac{5h_{3,\phi}(\phi,\sigma)}{4\phi h_3(\phi,\sigma)} - \frac{3h_{3,\phi}(\phi,\sigma)}{4\left(\phi - \sigma\right) 
h_3(\phi,\sigma)} - \frac{h_{3,\phi\phi}(\phi,\sigma)}{2h_3(\phi,\sigma)} + \left(\frac{h_{3,\phi}(\phi,\sigma)}{2h_3(\phi,\sigma)} \right)^2 \right]\, , \nonumber \\
V(\phi,\sigma)=& \, \frac{1}{2\kappa^2 \phi} \left( 1 - \frac{\sigma}{\phi} \right) \frac{h_{3,\phi}(\phi,\sigma)}{h_3(\phi,\sigma)} \,,
\end{align}
Here $\sigma$ is identified with the radius of the horizon, $\sigma=r_\mathrm{H}$. 

One should note that $\e^{-2\lambda(r)}$ must vanish when $\e^{2\nu(r)}$ vanishes in order to avoid the curvature singularity. 
Both $\e^{-2\lambda(r)}$ and $\e^{2\nu(r)}$ vanish at the horizon, one can write the horizon radius by $m_\mathrm{eff}(r)$, 
\begin{align}
\label{horizon}
r_\mathrm{H}= 2G m_\mathrm{eff}(r_\mathrm{H}) \, .
\end{align}
As we find the Hawking temperature (\ref{dS6BB}), we now consider the temperature of the black hole. 
Near the horizon, we write the radial coordinate as $r \equiv r_\mathrm{H} + \delta r$. 
Then we obtain, 
\begin{align}
\e^{-2\lambda} \sim \frac{C\left( r_\mathrm{H} \right) \left( r - r_\mathrm{H} \right) }{r_\mathrm{H}} \, , \quad 
\e^{2\nu} \sim \frac{C\left( r_\mathrm{H} \right) \left( r - r_\mathrm{H} \right) }{h_3 \left( r_\mathrm{H} \right) r_\mathrm{H}} \, .
\end{align} 
Here $C\left( r_\mathrm{H} \right)\equiv 1 - m'\left( r_\mathrm{H} \right)$. 
By a Wick rotation, $t\to i\tau$, the line element~(\ref{metric}) near the horizon behaves as 
\begin{align}
\label{TH1}
ds^2 \sim \frac{C\left( r_\mathrm{H} \right) \delta r}{h_3 \left( r_\mathrm{H} \right) r_\mathrm{H}} d\tau^2 
+ \frac{r_\mathrm{H}}{C\left( r_\mathrm{H} \right) \delta r} d(\delta r)^2 + r_\mathrm{H}^2 \, d\Omega_{(2)}^2 \, .
\end{align}
By using a new radial coordinate $\rho$ defined by 
$d\rho = d \left( \delta r\right) \sqrt{ \frac{r_\mathrm{H}}{C\left( r_\mathrm{H} \right) \delta r}}$, which gives, 
\begin{align}
\label{TH2}
\rho = 2 \sqrt{ \frac{r_\mathrm{H} \delta r}{C\left( r_\mathrm{H} \right)}} \quad
\mbox{or} \quad \delta r= \frac{C\left( r_\mathrm{H} \right) \rho^2}{4 r_\mathrm{H}} \, ,
\end{align}
we rewrite line element~(\ref{TH1}) as
\begin{align}
\label{TH3}
ds^2 \simeq \frac{C\left( r_\mathrm{H} \right)^2}{ 4 h_3 \left( r_\mathrm{H} \right) r_\mathrm{H}^2} \rho^2d\tau^2
+ d\rho^2 + r_\mathrm{H}^2 d\Omega_{(2)}^2 \,.
\end{align}
In order to avoid conical singularities near $\rho = 0$ in the Euclidean space, 
we need to impose the periodicity of the Euclidean time coordinate $\tau$, 
\begin{align}
\label{TH4}
\frac{C\left( r_\mathrm{H} \right) \tau}{ 2 r_\mathrm{H} \sqrt{h_3 \left( r_\mathrm{H} \right)}} 
\simeq \frac{C\left( r_\mathrm{H} \right) \tau}{ 2 \, r_\mathrm{H} \sqrt{h_3 \left( r_\mathrm{H} \right)}} + 2 \pi \, .
\end{align}
Because the period of the Euclidean time corresponds to the temperature $T$, we find 
\begin{align}
\label{TH6}
T = \frac{C\left( r_\mathrm{H} \right)}{4\pi r_\mathrm{H}\sqrt{h_3 \left( r_\mathrm{H} \right)}} 
= \frac{C\left( r_\mathrm{H} \right)}{8\pi G m_\mathrm{eff} \left( r_\mathrm{H} \right) \sqrt{h_3 \left( r_\mathrm{H} \right)} } 
= \frac{C\left( r_\mathrm{H} \right) T_\mathrm{H}}{\sqrt{h_3 \left( r_\mathrm{H} \right)}} \, .
\end{align}
Here the Hawking temperature $T_\mathrm{H}$ is now given by, 
\begin{align}
T_\mathrm{H}\equiv \frac{1}{8\pi G m_\mathrm{eff} \left( r_\mathrm{H} 
\right)} \,.
\end{align}
Therefore we find the temperature $T$ deviates from the Hawking temperature by the factor 
$\frac{C \left( r_\mathrm{H} \right)}{\sqrt{h_3 \left( r_\mathrm{H} \right)}}$, which cannot be absorbed by rescaling time. 

By the analogy of the thermodynamical relation $dE= T d\mathcal{S}$, we define the entropy proper to the black hole.
\begin{align}
\label{entropy2}
d\mathcal{S}_\mathrm{bh} = \frac{d m_\mathrm{eff}\left( r_\mathrm{H} 
\right)}{T_\mathrm{H}}\,.
\end{align}
By integrating (\ref{entropy2}), we obtain, 
\begin{align}
\label{entropy3B}
\mathcal{S}_\mathrm{bh} = \int \frac{d m_\mathrm{eff} \left( r_\mathrm{H} 
\right)}{T}\, .
\end{align}
We now consider the possibility that $\mathcal{S}_\mathrm{bh}$ could be different from the Bekenstein-Hawking entropy. 

By solving the field equations of a certain gravitational theory, there appear several constants of the integration, $c_i$ $\left(i=1, \, \cdots \, , N \right)$. 
For example, in general relativity, the mass $M$ of the Schwarzschild black hole (\ref{SchwrzP}) appears as an integration constant. 
Both the mass $M$ and charge $Q$ in the Reissner-Nordstr\"{o}m black hole (\ref{RNmetric}) are also constants of the integration. 
The horizon radius $r_\mathrm{H}$ could be given by a function of $c_i$ as in the usual Schwarzschild black hole, where. we find $r_\mathrm{H}=2GM$ as 
a function of the integration constant $M$. 
Other quantities could be also obtained as functions of $c_i$, such as $ h_3\left(r=r_\mathrm{H} \left(c_i\right); c_i\right)$, etc. 
We may also assume that the constants $c_i$'s are parametrised using a single parameter $\xi$, $c_i=c_i(\xi)$ as mentioned before Eq.~(\ref{SgRN}) 
in the case of the Reissner-Nordstr\"{o}m black hole. 

Eq.~(\ref{TH6}) can be used to rewrite Eq.~(\ref{entropy3B}) in the following form 
\begin{align}
\label{entropy3c}
\mathcal{S}_\mathrm{bh} = \frac{1}{2G} \int d\xi \frac{\left[ 4\pi r_\mathrm{H} \left( c_i \left(\xi\right) \right) 
\sqrt{ h_3\left(r=r_\mathrm{H} \left(c_i\left(\xi\right) \right); c_i\left(\xi\right)\right) } \right]}
{1 - \left. \frac{\partial m\left(r; c_i\left(\xi\right) \right)}{\partial r}\right|_{r= r_\mathrm{H} \left( c_i \left(\xi\right) \right)}}
\sum_{i=1}^N \frac{\partial r_\mathrm{H} \left(c_i\right)}{\partial c_i} \frac{\partial c_i}{\partial \xi}\, .
\end{align}
By choosing $\xi=r_\mathrm{H}$, Eq.~(\ref{entropy3c}) is simplified to be, 
\begin{align}
\label{entropy3d}
\mathcal{S}_\mathrm{bh} = \frac{1}{2G} \int_0^{r_\mathrm{H}} d\xi \, 
\frac{\left(4\pi \xi \sqrt{ h_3\left(r=\xi; c_i\left(\xi\right)\right) } 
\right) }{1 - \left. \frac{\partial m\left(r; c_i\left(\xi\right) 
\right)}{\partial r}\right|_{r= \xi}} \, .
\end{align}
Here the constant of the integration is fixed by using the condition $\mathcal{S}_\mathrm{bh}=0$ at $r_\mathrm{H} =0$. 
In the case of the Schwarzschild black hole, where $h_3(x)=1$, $m=M=\mbox{const.}$, 
Bekenstein-Hawking entropy~(\ref{S3}) is reproduced. 
In general, however, if $h_3\left( r \to r_\mathrm{H} \right)$ non-trivially contribution to the entropy, 
$\mathcal{S}_\mathrm{bh}$ may be different from the Bekenstein-Hawking entropy $\mathcal{S}_\mathrm{BH}$. 

In fact, Eq.~(\ref{entropy3d}) gives, 
\begin{align}
\label{entropy5}
\frac{h_3\left(r=r_\mathrm{H}; c_i\left(r_\mathrm{H}\right)\right)}
{\left(1 - \left. \frac{\partial m\left(r; c_i\left(r_\mathrm{H}\right) \right)}{\partial r}\right|_{r= r_\mathrm{H}}\right)^2}
= 16G^2 \Big[ \mathcal{S}_\mathrm{bh}' \left(A\right) \Big]^2 \, .
\end{align}
Therefore for certain expressions of the general entropies, we find the corresponding form of 
\begin{align}
\label{Theta}
\Theta \equiv \frac{h_3\left(r=r_\mathrm{H}; c_i\left(r_\mathrm{H}\right)\right) } 
{\left(1 - \left. \frac{\partial m\left(r; c_i\left(r_\mathrm{H}\right) 
\right)}{\partial r}\right|_{r= r_\mathrm{H}}\right)^2} \, .
\end{align}
For example, in the case of the R{\'e}nyi entropy~(\ref{RS1}), we obtain 
\begin{align}
\label{entropy8}
\Theta = \frac{1}{ \left( 1 + \frac{\pi \alpha {r_\mathrm{H}}^2}{G} \right)^2} \, ,
\end{align}
and for the Tsallis entropy (\ref{TS1}), Eq.~(\ref{entropy5}) becomes 
\begin{align}
\label{entropy6}
\Theta = \delta^2 \left( \frac{4\pi {r_\mathrm{H}}^2}{A_0} \right)^{2\left(\delta - 1 \right)} \, .
\end{align}
Furthermore for the three-parameter generalised entropy $\mathcal{S}_3$ in (\ref{intro-3}), we find 
\begin{align}
\label{general5h2}
\Theta = \frac{\alpha^2}{\gamma^2} \left[ 1 + \left(\frac{\pi\alpha 
{r_\mathrm{H}}^2}{\beta G} \right) \right]^{2\beta - 2} \,.
\end{align}
and the six-parameter entropy $\mathcal{S}_6$ in (\ref{intro-2}) yields
\begin{align}
\label{general1h2}
\Theta = \frac{1}{\left(\alpha_+ + \alpha_-\right)^2}
\, & \left[ \alpha_+ \gamma_+ \left(\frac{\pi {r_\mathrm{H}}^2}{G} 
\right)^{\gamma_+ -1}\left( 1 + \frac{\alpha_+}{\beta_+} 
\left(\frac{\pi {r_\mathrm{H}}^2}{G} \right)^{\gamma_+} \right)^{\beta_+ - 1} \right. \nonumber \\
&\, \quad \left. + \alpha_- \gamma_- \left(\frac{\pi {r_\mathrm{H}}^2}{G} 
\right)^{\gamma_- -1}\left( 1 + \frac{\alpha_-}{\beta_-} 
\left(\frac{\pi {r_\mathrm{H}}^2}{G} \right)^{\gamma_-} \right)^{-\beta_- - 1} \right]^2 \, .
\end{align}
Even for the four-parameter one $\mathcal{S}_4$ in (\ref{intro-1}), the five-parameter one $\mathcal{S}_5$ in (\ref{intro-4}), 
we can find the corresponding quantity $\Theta$. 

Application of the alternative entropies to the Bekenstein-Hawking entropy to black holes 
lead to inconsistencies in the thermodynamics as we discussed but 
the inconsistencies might be avoided for non-Schwarzschild black holes in modified gravity 
if the horizon radius and therefore the area appearing in Bekenstein's area law are modified as 
we have shown. 
Hence, the consistency of new entropy proposals with Hawking temperature and area law 
could be possible for the above black holes as it is shown in this section. 

\section{Photon Sphere and Black Hole Shadow}\label{generalshadow}

Recently, there has been much interest in BH shadow. 
Let us briefly discuss this topic here in relation to different BH thermodynamics.
A photon sphere is the set of the circular orbit of the photon. 
The radius $r_\mathrm{ph}$ of the photon sphere gives the radius $r_\mathrm{sh}$ of the black hole shadow as follows, 
\begin{align}
\label{shph}
r_\mathrm{sh}=\left. r\e^{-\nu(r)} \right|_{r=r_\mathrm{ph}}\, .
\end{align}
The orbit of the photon is governed by the following Lagrangian, 
\begin{align}
\label{ph1g}
\mathcal{L}= \frac{1}{2} g_{\mu\nu} \dot q^\mu \dot q^\nu = \frac{1}{2} \left( - \e^{2\nu} {\dot t}^2 
+ \e^{2\lambda} {\dot r}^2 + r^2 {\dot\theta}^2 + r^2 \sin^2 \theta {\dot\phi}^2 \right) \, .
\end{align}
Here the ``dot'' or ``$\dot\ $'' expresses the derivative with respect to the affine parameter. 
The fact that the geodesic of the photon is null tells $\mathcal{L}=0$. 
We find the conserved quantities corresponding to energy $E$ and angular momentum $L$ 
because there are no the explicit dependences on $t$ and $\phi$ in the Lagrangian $\mathcal{L}$, 
\begin{align}
\label{phEgMg}
E \equiv \frac{\partial \mathcal{L}}{\partial \dot t} = - \e^{2\nu} \dot t \, , \quad 
L \equiv \frac{\partial V}{\partial\dot\phi}= r^2 \sin^2 \theta \dot\phi \, . 
\end{align}
The total energy $\mathcal{E}$ of the system should be also conserved and given by, 
\begin{align}
\label{totalEg}
\mathcal{E} \equiv \mathcal{L} - \dot t \frac{\partial \mathcal{L}}{\partial \dot t} - \dot r \frac{\partial \mathcal{L}}{\partial \dot r} 
 - \dot\theta \frac{\partial \mathcal{L}}{\partial \dot\theta} - \dot\phi \frac{\partial \mathcal{L}}{\partial \dot\phi} = \mathcal{L} \, , 
\end{align}
We should note that $\mathcal{E}=\mathcal{L}$ vanishes identically $\mathcal{E}=\mathcal{L}=0$ for the null geodesic. 

Without any loss of generality, we can choose the coordinate system where the orbit of the photon is on the equatorial plane with $\theta=\frac{\pi}{2}$. 
For the coordinate choice, the condition $\mathcal{E}=\mathcal{L}=0$ can be written as 
\begin{align}
\label{geo1g}
0= - \frac{E^2}{2} \e^{-2 \left( \nu + \lambda\right)} + \frac{1}{2} {\dot r}^2 + \frac{L^2 \e^{- 2\lambda}}{2r^2} \, ,
\end{align}
We write this system in an analogous way to the classical dynamical system with potential $W(r)$, 
\begin{align}
\label{geo2g}
0 =\frac{1}{2} {\dot r}^2 + W(r)\, , \quad W(r) \equiv \frac{L^2 \e^{- 2\lambda}}{2r^2} - \frac{E^2}{2} \e^{-2 \left( \nu + \lambda\right)}\, .
\end{align}
Because the radius of the circular orbit is defined by $\dot r=0$, 
the radius is given by solving $W(r)= W'(r)=0$ by using the analogy with classical mechanics. 
For the Schwarzschild spacetime, we find $r_\mathrm{ph}=3M$ and $r_\mathrm{sh}=3\sqrt{3} M$. 

In the model~(\ref{exam1}) with (\ref{h2}), we find 
\begin{align}
\label{Wex}
W(r) = \frac{L^2 \left( r - r_H\right)}{2r^2 \left(r + 2G M \left( r_\mathrm{H} \right) - r_\mathrm{H}\right)} - \frac{E^2}{2} \, ,
\end{align}
which gives 
\begin{align}
\label{Wexd}
W'(r) = - \frac{2L^2\left( r^2 + G M \left( r_\mathrm{H} \right) r - 2 r_\mathrm{H} r - 2 G M \left( r_\mathrm{H} \right) r_\mathrm{H} + {r_\mathrm{H}}^2 
\right)}{2r^3 \left(r + 2G M \left( r_\mathrm{H} \right) - r_\mathrm{H}\right)^2} \, ,
\end{align}
which gives 
\begin{align}
\label{ph2}
r = r_\mathrm{ph} \equiv 
\frac{1}{2} \left[ - G M \left( r_\mathrm{H} \right) + 2 r_\mathrm{H} \pm \sqrt{ G^2 M \left( r_\mathrm{H} \right)^2 
+ 4 G M \left( r_\mathrm{H} \right) r_\mathrm{H} } \right] \, .
\end{align}
In the Schwarzschild black hole case, $2G M \left( r_\mathrm{H} \right) = r_\mathrm{H}$, the above expression gives 
$r=0,\, 3 G M \left( r_\mathrm{H} \right)$. 
The case of $r=0$ is unphysical because the origin is inside the horizon. 
The second case $r=3 G M \left( r_\mathrm{H} \right)$ gives the standard result. 
In general, the minus signature in the front of the square root in (\ref{ph2}) gives the radius smaller than the horizon radius and 
therefore we choose the plus signature in (\ref{ph2}). 

Then Eq.~(\ref{shph}) gives the radius $r_\mathrm{sh}$ of the black hole shadow, 
\begin{align}
\label{shphex}
r_\mathrm{sh}
= \frac{2 G M \left( r_\mathrm{H} \right)}{8} \left( 3 + \sqrt{ 1 + 8\eta}\right)^\frac{3}{2} \left( - 1 + \sqrt{ 1 + 8\eta }\right)^\frac{1}{2} \, .
\end{align}
Here 
\begin{align}
\label{eta}
\eta \equiv \frac{r_\mathrm{H}}{2 G M \left( r_\mathrm{H} \right)}\, ,
\end{align}
which is a unity for the Schwarzschild black hole. 

We now compare the obtained result with the observation. 
For M87$^*$, the constraint for the radius is given by $\frac{2r_\mathrm{sh}}{GM} \sim 11.0\pm 1.5$~\cite{Bambi:2019tjh} 
or $\frac{r_\mathrm{sh}}{GM} \sim 5.5\pm 0.8$ and 
For Sgr A$^*$, we have $4.21\lesssim \frac{r_\mathrm{sh}}{GM} \lesssim 5.56$~\cite{Vagnozzi:2022moj}. 
By using the parameter $\eta$ (\ref{eta}), the constraint from M87$^*$ is rewritten as $0.86<\eta<1.33$ and 
Sgr A$^*$ as $0.73<\eta<1.11$. 
Therefore, the results are consistent with the Schwarzschild black hole, where $\eta=1$. 

If by future observations, we find $\eta$ could not be unity, the black hole is different from the Schwarzschild one and thermodynamics could be different from that of 
the Schwarzschild black hole. 
If we also obtain more information like the ADM mass of the black hole, we may obtain some clues to consider what kind of thermodynamics the black hole obeys. 
Especially if we obtain the information from several black holes, we may find more universal thermodynamics which governs the black holes. 

\section{Microcanonical and Canonical Description for Generalised Entropy}\label{SecVII}

 From the viewpoint of quantum gravity, the microscopic understanding of generalised entropy could be important and suggestive. 
In this section, based on \cite{Nojiri:2023ikl}, we consider the origins of various entropies in microscopic particle descriptions of the thermodynamical system. 
Note that basically, microscopic description gives some particle system which obeys the corresponding statistics (entropy). 
As we have entropy which depends on several parameters, we can eventually propose novel, not yet discovered information and statistical systems which obey these entropies.

\subsection{Microcanonical Description}\label{Sec2}

In this subsection, by using the microcanonical ensemble in thermodynamics, we consider how various generalised entropies appear 
in the isolated system with fixed energy $E$.

The standard Gibbs entropy is expressed as, 
\begin{align}
\label{S1B}
\mathcal{S}(E) = - \sum_{i=1}^{W(E)} P_i (E) \ln P_i (E) \, .
\end{align}
We choose the Boltzmann constant to be unity. 
Under the assumption the number of states with a fixed energy $E$ is $W(E)$ and a probability realising the $i$-th state with the energy $E$ is denoted by $P_i(E) $. 
Therefore we obtain, 
\begin{align}
\label{S2BB}
\sum_{i=1}^{W(E)} P_i (E) = 1 \, .
\end{align}
A generalization of the entropy with a parameter $\delta $ is proposed in \cite{Tsallis:1987eu} by Tsallis, 
\begin{align}
\label{S3B}
\mathcal{S}_\delta (E) \equiv \frac{ 1 - \sum_{i=1}^{W(E)} \left( P_i (E) \right)^\delta }{\delta -1} 
= \sum_{i=1}^{W(E)} \frac{P_i (E) \left( 1 - \left(P_i (E) \right)^{\delta -1} \right)}{\delta -1} \, .
\end{align}
In the limit of $\delta \to 1$, $\mathcal{S}_\delta (E) $ reduces to the standard expression in (\ref{S1B}). 

A further generalisation is given by the following expression, 
\begin{align}
\label{S4}
\mathcal{S}(E) = \sum_{i=1}^{W(E)} s_i \left( P_i (E) \right) \, .
\end{align}
We may regard $\mathcal{S}(E)$ as a function of $P_i$ and consider the maximum of $\mathcal{S}(E)$ in (\ref{S4}) 
under the constraint (\ref{S2BB}), which is nothing but the thermal equilibrium. 
Then we obtain an expression of the generalised entropy in the thermal equilibrium in the following form 
(see \cite{Nojiri:2023ikl} for more detailed calculations), 
\begin{align}
\label{S9}
\mathcal{S}(E) = W (E) s \left( \frac{1}{W(E)} \right) \, .
\end{align}
In the microcanonical approach, we define the temperature $T$ by 
\begin{align}
\label{T}
\frac{1}{T} \equiv \frac{dS(E)}{dE}\, .
\end{align}
This expression corresponds to (\ref{S1B}). 

We may consider the continuous phase space of $N$ particles $\left(q^i, p_i \right)$ $\left( i=1,2,\cdots, N \right)$ instead of considering the discrete states, 
which may be regarded with the limit of $W(E)\to \infty$. 
In this limit, Eq.~(\ref{S4}) has the following forms, 
\begin{align}
\label{S10}
S= \int_E \prod_{i=1}^N \left( \frac{dq^i dp_i}{\hbar} \right) s \left( q^k, p_k, P\left( q^k, p_k, E \right) \right)\, . 
\end{align}
We should note that $s$ may generally depend on $q^i$ and $p_i$ explicitly. 
In (\ref{S10}), \linebreak  $\int_E \prod_{i=1}^N \left( \frac{dq^i dp_i}{\hbar} \right) \cdots $ expresses the integration of the phase space for fixed energy $E$. 

In general, the function $s$ includes a finite or infinite number of parameters, $\left\{\alpha_n\right\}$, $n=1,2,\cdots$, 
$s=s \left( \left\{\alpha_n\right\}; q^k, p_k, P\left( q^k, p_k, E \right) \right)$, 
In a limit of the parameters, $\left\{\alpha_n\right\}$, $s$ may reduce to that in the Gibbs entropy (\ref{S1B}). 
As mentioned in Subsection~\ref{subge}, we impose the following conditions for $s$ with the parameters $\left\{\alpha_n\right\}$, 
\begin{enumerate}
\item Generalised third law: The generalised entropy vanishes when temperature $T$ vanishes as in the case of the Gibbs entropy (\ref{S1B}). 
Note, however, Bekenstein-Hawking entropy $\mathcal{S}$ (\ref{S3}) for black hole diverges when Hawking temperature $T_\mathrm{H}$ vanishes and 
$\mathcal{S}$ vanishes when $T_\mathrm{H}\to \infty$. 
\item Monotonicity: The generalised entropy is a monotonically increasing function of Gibbs entropy (\ref{S1B}). 
\item Positivity: The generalised entropy should be positive, as the number of states is greater than unity.
\item Gibbs entropy limit: The generalised entropy reduces to the Gibbs entropy (\ref{S1B}) in an appropriate limit of the parameters $\left\{\alpha_n\right\}$. 
\end{enumerate}
In standard thermodynamics, the following zeroth law must be also imposed, 
\begin{itemize}
\item When two systems denoted by $A$ and $B$ are in thermal equilibrium with a third system denoted by $C$, the system $A$ is also in equilibrium with the system $B$. 
\end{itemize}
The zeroth law does not hold in the case of non-extensive entropies like the Tsallis entropy \cite{Nauenberg:2002azf}. 
This tells the generalised entropies do not always satisfy the zeroth law. 

As we obtain (\ref{S9}) (see \cite{Nojiri:2023ikl} for more detailed calculations), we find 
\begin{align}
\label{S16}
S= V_\mathrm{phase} s \left( \frac{1}{V_\mathrm{phase}} \right) \, .
\end{align}
Here $V_\mathrm{phase}$ is the volume of the phase space, which can be finite because the energy $E$ is fixed.
By the choice of $s$, we obtain several kinds of entropy. 

Just for a simple example, we may consider one non-relativistic particle with mass $m$ moving on the two-dimensional space with the area $A$. 
Because the energy $E$ is fixed and given by 
\begin{align}
\label{S17}
E=\frac{{p_x}^2 + {p_y}^2}{2m}\, , 
\end{align}
the volume of the momentum space is equal to the area of a two-dimensional sphere with the radius $\sqrt{2m E}$, $4\pi \left( \sqrt{2m E} \right)^2 = 8\pi m E$. 
Therefore we obtain 
\begin{align}
\label{S18}
V_\mathrm{phase}= \frac{8\pi m E A}{\hbar^3}\, .
\end{align}
By the choice of $s(\xi)=-\xi\ln \xi$, the standard expression of the Gibbs entropy, denoted by $\mathcal{S}_0$ is obtained, which we now denote $S_0$, 
\begin{align}
\label{S19}
\mathcal{S}_0 = \ln V_\mathrm{phase} = k \ln \left( \frac{8\pi m E A}{\hbar^3} \right)\, .
\end{align}
On the other hand, if $s(\xi)$ is given by 
\begin{align}
\label{S20}
s(\xi) = \frac\xi{\gamma} \left[ \left( 1 - \frac{\alpha}{\delta} \ln \xi \right)^\delta - 1 \right] \, ,
\end{align}
with positive dimensionless parameters $ \left( \alpha, \gamma, \delta \right)$, 
we obtain an expression similar to the three-parameter entropy (\ref{intro-3}) in \cite{Nojiri:2022aof}. 
On the other hand, if $s(\xi)$ is given by 
\begin{align}
\label{S22}
s(\xi) = \frac\xi{\gamma}\left[\left(1 - \frac{\alpha_+}{\delta} \ln \xi \right)^\delta 
 - \left(1 - \frac{\alpha_-}{\delta} \ln \xi \right)^{-\delta}\right] \,,
\end{align}
we obtain an expression corresponding to a four-parameters generalised entropy (\ref{intro-1}) proposed in \cite{Nojiri:2022dkr}. 
It is straightforward to find $s(\xi)$ corresponding to other versions of generalised entropy. 

In the case of non-extensive systems, such as gravitational or electromagnetic ones, 
the standard Gibbs additive entropy (\ref{S1B}) should be replaced by the non-extensive Tsallis entropy~\cite{Tsallis:1987eu}. 
The non-extensive entropy tells that the numbers of the states show the running behaviour by the change of the energy scale, 
as in the renormalisation group of quantum field theory. 
Because the entropy corresponds to the physical degrees of freedom of a system, the renormalisation group 
of a quantum theory implies that the degrees of freedom depend on the energy scale. 
In the low-energy regime, massive modes decouple, and therefore the degrees of freedom decrease. 
In the case of gravity, if the space-time fluctuations become large in the ultraviolet regime, the
degrees of freedom might increase. 
On the other hand, if gravity becomes topological, the degrees of freedom decrease. 
The latter situation is consistent with holography. 
This could suggest that the generalised entropy might also appear by reflecting the quantum structure of gravity. 

\subsection{Canonical Description}\label{Sec3}

We now consider the canonical ensemble in thermodynamics, where the system is in equilibrium with the heat bath with temperature $T$. 
Even for the canonical ensemble, various versions of entropy appear to originate from the integration measure in the phase space. 

The partition function of $N$ particles is defined by 
\begin{align}
\label{GS1}
Z(\tilde\beta) = \int \prod_{i=1}^N \left( \frac{dq^i dp_i}{\hbar} \right) \e^{- \tilde\beta H \left( q^i, p_i \right)}\, .
\end{align}
Here $q^i$ and $p_i$ are the coordinates of the position and the momenta for the $i$-th particle, respectively, as in the last subsection. 
We define $\tilde\beta$ as susual, $\tilde\beta \equiv \frac{1}{T}$. 
The reason why we use the measure $ \prod_{i=1}^N \left( \frac{dq^i dp_i}{\hbar} \right)$ is because it is invariant under the canonical transformation 
in classical mechanics. 
We should note, however, that in quantum mechanics, only the cartesian coordinates have a special meaning. 

More in general, instead of $\prod_{i=1}^N \left( \frac{dq^i dp_i}{\hbar} \right)$, we may consider measure given by 
$\e^{-M\left(q^i,p_i\right)}\prod_{i=1}^N \left( \frac{dq^i dp_i}{\hbar} \right)$ and we may define the partition function as follows, 
\begin{align}
\label{GS8}
Z(\tilde\beta) = \int \prod_{i=1}^N \left( \frac{dq^i dp_i}{\hbar} \right) \e^{- \tilde\beta H \left( q^i, p_i \right)- M\left(q^i,p_i\right)}\, .
\end{align}
In the situation that we confine the particles in the box with edge length $L$, $M\left(q^i,p_i\right)$ is given by 
\begin{align}
\label{GS9}
\e^{-M\left(q^i,p_i\right)} = \prod_{i=1}^N \theta \left( q^i \right) \theta \left( L - q^i \right) \, , \quad \mbox{or} \quad 
M = - \sum_{i=1}^N \left( \ln \theta \left( q^i \right) + \ln \theta \left( L - q^i \right) \right) \, .
\end{align}
Here $\theta(\xi)$ is the usual Heaviside step function, 
\begin{align}
\label{CS10}
\theta(\xi) = \left\{ \begin{array}{cc} 1 & \mbox{when}\ \xi \geq 0 \\
0 & \mbox{when}\ \xi<0 
\end{array} \right. \, .
\end{align}
We may consider the following model as an example, 
\begin{align}
\label{CS11}
\left(q^i\right) =&\, \left( x,y,z\right)\, , \quad 
H=\frac{1}{2m} \left( p_x^2 + p_y^2 + p_z^2 \right) \, , \nonumber \\
\e^{-M}=&\, { 4\pi R^2} \delta\left( R^2 - x^2 - y^2 -z^2 \right) \e^{-X\left( 4\pi \left( x^2 + y^2 + z^2 \right)\right)}\, .
\end{align}
Here $X$ is an adequate function. 
After the integration in the phase space, we obtain, 
\begin{align}
\label{CS12}
Z(\tilde\beta)= \frac{ 8\pi^2 R^3}{\hbar^3} \left( \frac{2m \pi}{\tilde\beta} \right)^\frac{3}{2} \e^{-X\left(4\pi R^2 \right)}\, ,
\end{align}
which gives the following free energy $F(\tilde\beta)$, 
\begin{align}
\label{CS13}
F(\tilde\beta) = - \frac{1}{\tilde\beta} \ln Z(\tilde\beta) 
= - \frac{1}{\tilde\beta} \left( \ln \left( \frac{8\pi^2 R^3}{\hbar^3} \left( \frac{2m \pi}{\tilde\beta}\right)^\frac{3}{2} \right)
 - X\left(4\pi R^2 \right) \right) \, .
\end{align}
This expression give the following thermodynamical energy $E\left( \tilde\beta \right)$, 
\begin{align}
\label{CS14}
E(\tilde\beta) = F(\tilde\beta) + \tilde\beta \frac{\partial F(\tilde\beta)}{\partial \tilde\beta} = \frac{3}{2\tilde\beta}\, ,
\end{align} 
and the entropy $\mathcal{S}$ 
\begin{align}
\label{CS15}
\mathcal{S}= \tilde\beta \left( E-F \right) = \left\{ \frac{3}{2} 
+ \ln \left( \frac{8\pi^2 R^3}{\hbar^3} \left( \frac{2m \pi}{\tilde\beta} \right)^\frac{3}{2} \right) - X\left(4\pi R^2 \right) \right\} \, .
\end{align}
If we assume $X(\xi)$ is given by $X(\xi)=- \frac{\xi}{4G}$ with Newton's gravitational constant $G$, 
the last term in (\ref{CS15}) may dominate for large $R$, which results in Bekenstein-Hawking entropy, 
\begin{align}
\label{CS16}
\mathcal{S} =\frac{A}{4G}\, , \quad A\equiv 4\pi R^2\, .
\end{align}
On the other hand, if we choose $X(\xi)$ by $X(\xi)=- \frac{A_0}{4G}\left( \frac{\xi}{A_0}\right)^\delta$, Tsallis entropy in (\ref{TS1}) 
can be obtained, 
\begin{align}
\label{CS17}
\mathcal{S}\to \frac{A_0}{4G}\left( \frac{A}{A_0}\right)^\delta\, .
\end{align}
The function $M\left(q^i,p_i\right)$ appearing in the measure should be given by the properties of the corresponding physical system 
but we can find the measure which gives the corresponding kind of (generalised) entropy. 

In the case of R{\'e}nyi entropy in (\ref{RS1}), we find $X(\xi)=-\frac{1}{\alpha} \ln \left( 1 + \frac{\alpha \xi}{4G} \right)$. 
For the three-parameter entropy (\ref{intro-3}), we obtain 
\begin{align}
\label{CS18}
X(\xi) = - \frac{1}{\gamma} \left[ \left( 1 + \frac{\alpha {\xi}}{4G \delta} \right)^\delta - 1 \right] \,,
\end{align}
Further, a four-parameters generalised entropy (\ref{intro-1}) is given by 
\begin{align}
\label{CS19}
X(\xi) = - \frac{1}{\gamma}\left[\left(1 + \frac{\alpha_+ \xi}{4G\delta}\right)^\delta
 - \left(1 + \frac{\alpha_- \xi}{4G \delta} \right)^{-\delta}\right] \, .
\end{align}
Thus we have shown that the function $X(x)$ corresponding to the generalised entropy can be always found. 

The general measure may originate from the modification of the commutation relation 
$\left[ q^i, p_j \right]=i \hbar \delta^i_{\ j}$. 
We consider the following commutation relation 
(here we write the reduced Planck constant or Dirac's constant $\hbar$ explicitly), 
\begin{align}
\label{CS24}
\left[ q^i, p_i \right] = i\hbar \e^{M^i_{\ j}(q^k,p_k)}\, ,
\end{align}
which induces the metric in the phase space as follows, 
\begin{align}
\label{CS25}
ds^2 = \sum_{i,j=1}^N g^j_{\ i}dq^i dp_j \, , \quad g^j_{\ i} \equiv \left( L^{-1} \right)^j_{\ i} \, , L^i_{\ j}\equiv \e^{M^i_{\ j}(q^k,p_k)} \, .
\end{align}
Here $\left( L^{-1} \right)^j_{\ i}$ is the inverse matrix of $L^i_{\ j}$ when $L^i_{\ j}$ is regarded as $N\times N$ matrix, 
$\sum_{k=1}^N L^i_{\ k} \left( L^{-1} \right)^k_{\ j} = \delta^i_{\ j}$. 
The metric $g^j_{\ i}$ gives the following volume form, 
\begin{align}
\label{CS26}
dV = \det \left( g^j_{\ i} \right) \prod_{i=1}^N \left( dq^i dp_i \right)\, .
\end{align}
Due to the symplectic structure of the phase space, $\det \left( g^j_{\ i} \right)$ is a Pfaffian. 
In the case that $\e^{M^i_{\ j}(q^k,p_k)}$ is proportional to the unit matrix, $\e^{M^i_{\ j}(q^k,p_k)}=\e^{\frac{1}{N}M\left( q^i, p_i \right)}{\delta^i}_j$, 
$dV$ reduces to the previous expression of the general measure, 
\begin{align}
\label{CS26BBB}
dV = \hbar^N \e^{-M\left(q^i,p_i\right)}\prod_{i=1}^N \left( \frac{dq^i dp_i}{\hbar} \right) \, .
\end{align}
We should note that we cannot rewrite the metric in (\ref{CS25}) and the commutation relations in (\ref{CS24}) 
in a diagonal form like $\left[ Q^i, P_i \right] = i\hbar \delta^i_{\ j}$ 
by any redefinition of the variables $Q^i=Q^i\left( q^j, p_j \right)$, $P_i =P_i \left( q^j, p_j \right)$ if there is a non-trivial curvature given by the metric in (\ref{CS25}). 

For the three-parameter entropy (\ref{intro-3}), by using (\ref{CS11}) with (\ref{CS18}), 
we find Eq.~(\ref{CS24}) has the following form, 
\begin{align}
\label{CS27}
\left[ q^i, p_j \right] = \frac{i\hbar \e^{\frac{1}{\gamma} \left[ \left( 1 + \frac{\alpha { c^3} \pi \left( x^2 + y^2 + z^2 \right)}{{ \hbar} G \beta} \right)^\beta - 1 \right] }}
{{ 4\pi R^2} \delta\left( R^2 - x^2 - y^2 -z^2 \right)} \delta^i_{\ j} \, .
\end{align}
Here the inverse power of the delta function does not have a physical meaning but the delta function $\delta(x)$ can be defined by, 
\begin{align}
\label{CS28}
\delta (x) \equiv \lim_{\lambda\to \infty} \sqrt{\frac{\lambda}{\pi}} \e^{-\lambda x^2}\, .
\end{align}
This suggests that instead of (\ref{CS27}), by choosing the parameter $\lambda$ sufficiently large, we replace the commutation relation in (\ref{CS27}) by, 
\begin{align}
\label{CS29}
\left[ q^i, p_j \right] =&\, \frac{i\hbar}{ 4\pi R^2} \sqrt{\frac{\pi}{\lambda}} \e^{\frac{1}{\gamma} \left[ \left( 1 + \frac{\alpha { c^3} 
\pi \left( x^2 + y^2 + z^2 \right)}{{ \hbar}G \beta} \right)^\beta - 1 \right] 
+ \lambda \left( R^2 - x^2 - y^2 -z^2 \right)^2 } \delta^i_{\ j} \, , \nonumber \\
i,j=&\, x,y,z, \quad \left(q^x, q^y, q^z \right)=\left(x,y,z \right)\, ,
\end{align}
which might be the origin of the three-parameter entropy (\ref{intro-3}). 

It is known that due to the generalised uncertainty principle based on the introduction of the minimal length 
\cite{Yoneya:1989ai}, the modification of the canonical commutation relations could be generated. 
The motivation of the minimal length comes from string theory where the minimum size of the fundamental string is finite. 

\section{Microscopic Interpretation of Generalised Entropy}\label{SecVIII}

Except the analogy of the Tsallis entropy in (\ref{TS1}) and the Barrow entropy~\cite{Barrow:2020tzx}, 
the generalised entropies $\mathcal{S}_\mathrm{g}$, which are functions of the Bekenstein-Hawking entropy $\mathcal{S}$, coincide 
with $\mathcal{S}$ in the limit of $\mathcal{S}\to 0$ and they have the McLaughlin expansion with respect to $\mathcal{S}$, 
\begin{align}
\label{McLaughlin}
\mathcal{S}_\mathrm{g} \sum_{n=0}^{\infty} \frac{f_\mathrm{g}^{(n)}}{n!}\mathcal{S}^n\, .
\end{align}
Here $f_\mathrm{g}^{(n)}$ is defined by $f_\mathrm{g}^{(n)} \equiv \left. \frac{\partial^n \mathcal{S}_\mathrm{g}}{\partial \mathcal{S}^n} \right|_{\mathcal{S}=0}$ 
and the functions of the parameters specifying the generalsed entropy $\mathcal{S}_\mathrm{g}$. 
The explicit forms of $f_\mathrm{g}^{(n)}$ for $\mathcal{S}_3$ in (\ref{intro-3}), the four-parameter one $\mathcal{S}_4$ in (\ref{intro-1}), 
the five-parameter one $\mathcal{S}_5$ in (\ref{intro-4}), and the six-parameter entropy $\mathcal{S}_6$ in (\ref{intro-2}), see \cite{Nojiri:2023bom}. 
In this section, based on \cite{Nojiri:2023bom}, we consider the generalised entropies in the microscopic viewpoint of the canonical 
and grand canonical ensembles. 

\subsection{Canonical Description}\label{sec-c}

In the canonical prescription, the phase space density of a thermodynamical system composed of $N$ particles is expressed as 
\begin{align}
\rho_\mathrm{c}\left(q^j,p_j\right) = \frac{\exp\left(-\tilde\beta H\right)}{Z(T,V,N)}\, ,
\label{canonical-3}
\end{align}
where $\tilde\beta = \frac{1}{T}$. 
Here $T$ is the temperature as before and we choose the Boltzmann constant as unity. 
The index $j$ runs from $j=1$ to $j=3N$ and $\left\{q^j,p_j\right\}$ are generalised coordinates and generalised momenta of the system, respectively. 
We denote the Hamiltonian of the system by $H(q^j,p_j)$ and
\begin{align}
Z(T,V,N) = \int \frac{d^{3N}qd^{3N}p}{h^{3N}}\e^{-\tilde\beta H}\, ,
\label{canonical-4}
\end{align}
is the partition function which depends on temperature ($T$), volume ($V$) and number of particles ($N$) of the system. 
In (\ref{canonical-4}), h is the Planck constant. 
The expression of the partition function $Z(T,V,N)$ (\ref{canonical-4}) tells that the total probability is surely unity, 
\begin{align}
\int \frac{d^{3N}q d^{3N}p}{h^{3N}} \rho_\mathrm{c}(q^j,p_j) = 1\, ,
\label{canonical-5}
\end{align}
which allows us to define the ensemble average of a general microscopic quantity $v(q^j,p_j)$ by,
\begin{align}
\left\langle v(q,p) \right\rangle = \int \frac{d^{3N}q d^{3N}p}{h^{3N}} v(q,p)\rho_\mathrm{c}(q,p)\, .
\label{canonical-6}
\end{align}
The Gibbs entropy which we denote by $\mathcal{S}_0$ corresponding to (\ref{S1}) 
is defined by 
\begin{align}
\mathcal{S}_0 = \left\langle - \ln{\rho_\mathrm{c}} \right\rangle= -\int \frac{d^{3N}q d^{3N}p}{h^{3N}} \rho_\mathrm{c}\ln \rho_\mathrm{c}
= \tilde\beta\left\langle H \right\rangle + \ln{Z}\, .
\label{canonical-7}
\end{align}
Here we have used Eqs.~(\ref{canonical-4}), (\ref{canonical-5}), and (\ref{canonical-6}). 
Similarly we obtain the ensemble average of $(-\ln{\rho_\mathrm{c}})^2$ as follows, 
\begin{align}
\left\langle \left(-\ln{\rho_\mathrm{c}}\right)^2 \right\rangle= \int \frac{d^{3N}q d^{3N}p}{h^{3N}} \rho_\mathrm{c}\left(\ln{\rho_\mathrm{c}}\right)^2 
= \tilde\beta^2\left\langle H^2 \right\rangle + 2 \tilde\beta \left\langle H \right\rangle\ln{Z} + \left(\ln{Z}\right)^2\, ,
\label{canonical-8}
\end{align}
which is rewritten as, 
\begin{align}
\left\langle \left(-\ln{\rho_\mathrm{c}}\right)^2 \right\rangle = {\mathcal{S}_0}^2 + \tilde\beta^2\sigma_2(H)\, .
\label{canonical-9}
\end{align}
Here e define $\sigma_2(H) = \left\langle H^2 \right\rangle - \left\langle H \right\rangle^2$. 
Therefore we obtain, 
\begin{align}
{\mathcal{S}_0}^2 = \left\langle \left(- \ln{\rho_\mathrm{c}}\right)^2\right\rangle - \tilde\beta^2\sigma_2(H)\, ,
\label{canonical-10}
\end{align}
that is, ${\mathcal{S}_0}^2$ is the sum of the ensemble average of $\left(-\ln{\rho_\mathrm{c}}\right)^2$ and a term including $\sigma_2(H)$. 
In the standard extensive thermodynamical system, $\frac{\sigma_2(H)}{\left\langle H \right\rangle}$ is proportional to $\frac{1}{\sqrt{N}}$, 
which goes to vanish in the thermodynamic limit, $N \rightarrow \infty$, 
and Eq.~(\ref{canonical-10}) reduces to the form ${\mathcal{S}_0}^2 = \left\langle\left(-\ln{\rho_\mathrm{c}}\right)^2\right\rangle$. 
However for non-extensive systems, $\frac{\sigma_2(H)}{\left\langle H \right\rangle}$ does not vanish even in the thermodynamic limit and 
${\mathcal{S}_0}^2$ includes the extra term as in the second term of Eq.~(\ref{canonical-10}). 

By the similar procedure, we obtain ${\mathcal{S}_0}^n$ for general positive integer $n$ as follows, 
\begin{align}
{\mathcal{S}_0}^n = \left\langle \left(- \ln{\rho_\mathrm{c}}\right)^n\right\rangle - \sum_{l=2}^n \frac{n!}{l!(n-l)!}\left(\tilde\beta\right)^l\sigma_l(H)\left(\ln{Z}\right)^{n-l}\, ,
\label{canonical-13} \, .
\end{align}
Here $\sigma_l (H) = \left\langle H^l \right\rangle - \left\langle H \right\rangle^l$. 
Because $\sigma_\mathrm{1}(H) = 0$, we can take sum in the second term of (\ref{canonical-13}) from $l=1$, $\sum_{l=2}^n \to \sum_{l=1}^n$. 

By using
\begin{align}
\left\langle H^l \right\rangle = \frac{1}{Z} \int \frac{d^{3N}q~d^{3N}p}{h^{3N}} \e^{-\tilde\beta H} H^l\, ,
\label{canonical-14}
\end{align}
we can express $\sigma_l(H)$ in terms of the partition function $Z$, as follows, 
\begin{align}
\sigma_l (H) = \left(-1\right)^l \left\{\frac{1}{Z}\frac{\partial^l Z}{\partial\tilde\beta^l} - \left(\frac{1}{Z}\frac{\partial Z}{\partial\tilde\beta}\right)^l\right\}\, .
\label{canonical-15}
\end{align}
By using the expression of ${\mathcal{S}_0}^n$ in (\ref{canonical-13}) with (\ref{canonical-15}), we define an entropy similar to the form of generalised entropy, as follows, 
\begin{align}
\mathrm{S}_\mathrm{can}=&\, \sum_{n=0}^{\infty} \frac{f_\mathrm{g}^{(n)}}{n!} {\mathcal{S}_0}^n \nonumber\\
=&\, \sum_{n=0}^\infty \frac{f_\mathrm{g}^{(n)}}{n!}\left\{\left\langle \left(- \ln{\rho_\mathrm{c}}\right)^n\right\rangle 
 - \sum_{l=2}^n\frac{n!}{l!(n-l)!}\left(\tilde\beta\right)^l\sigma_l (H)\left(\ln{Z}\right)^{n-l}\right\}\, .
\label{canonical-16}
\end{align}
Especially in the cases of the three-parameter entropy $\mathcal{S}_3$ in (\ref{intro-3}) and the four-parameter one $\mathcal{S}_4$ in (\ref{intro-1}), 
we obtain 
\begin{align}
\label{canonical-17}
\mathcal{S}_\mathrm{can\, 3}=&\, \sum_{n=0}^{\infty} \frac{f_3^{(n)}\left(\alpha,\delta,\gamma\right)}{n!}
\left\{\left\langle \left(-\ln{\rho_\mathrm{c}}\right)^n \right\rangle - \sum_{l=2}^{n}\frac{n!}{l!(n-l)!}\left(\tilde\beta\right)^l 
\sigma_l(H)\left(\ln{Z}\right)^{n-l}\right\}\nonumber\\
=&\, \frac{1}{\gamma}\left[\left(1 + \frac{\alpha}{\delta}~\mathcal{S}_0\right)^{\delta} - 1\right]\, , \\
\label{canonical-18}
S_\mathrm{can\, 4}=&\, \sum_{n=0}^{\infty} \frac{f_4^{(n)}\left(\alpha_{\pm},\delta,\gamma\right)}{n!}
\left\{\left\langle \left(-\ln{\rho_\mathrm{c}}\right)^n\right\rangle - \sum_{l=2}^{n}\frac{n!}{l!(n-l)!}\left(\tilde\beta\right)^l \sigma_l(H)\left(\ln{Z}\right)^{n-l}\right\}\nonumber\\
=&\, \frac{1}{\gamma}\left[\left(1 + \frac{\alpha_+}{\delta}~\mathcal{S}_0\right)^{\delta}
 - \left(1 + \frac{\alpha_-}{\delta}~\mathcal{S}_0\right)^{-\delta}\right]\, .
\end{align}
In this way, we may consider the cases of the five-parameter one $\mathcal{S}_5$ in (\ref{intro-4}), 
and the six-parameter entropy $\mathcal{S}_6$ in (\ref{intro-2}). 

\subsection{Grand Canonical Description}\label{sec-gc}

The grand canonical phase space density with chemical potential $\mu$ in addition to Hamiltonian $H$ is defined by,
\begin{align}
\rho_\mathrm{gc}\left(q^j,p_j,N\right) = \frac{\exp\left\{-\tilde\beta\left(H-\mu N\right)\right\}}{\mathcal{Z}(T,V,\mu)}\, .
\label{g-canonical-1}
\end{align}
Again, $j$ runs from $j=1$ to $j=3N$. 
Because the particle number $N$ in a grand canonical ensemble fluctuates, a single microstate is characterised by $\left\{q^j, p_j, N\right\}$. 
In (\ref{g-canonical-1}), $\mathcal{Z}$ is a grand canonical partition function given by, 
\begin{align}
\mathcal{Z}(T,V,\mu) = \sum_N \int \frac{d^{3N}q d^{3N}p}{h^{3N}}\e^{-\tilde\beta\left(H-\mu N\right)}\, .
\label{g-canonical-2}
\end{align}
Then the ensemble average of a microscopic quantity $v(q^j,p_j,N)$ in grand canonical description is given by,
\begin{align}
\left\langle v(q,p,N) \right\rangle = \sum_N \int \frac{d^{3N}q d^{3N}p}{h^{3N}} v(q,p,N)\rho_\mathrm{gc}(q,p,N)\, .
\label{g-canonical-3}
\end{align}

For grand canonical ensemble, the Gibbs entropy symbolized by $\mathcal{S}_0$ is defined by, 
\begin{align}
\mathcal{S}_0 = \left\langle -\ln{\rho_\mathrm{gc}} \right\rangle= -\sum_N \int \frac{d^{3N}q d^{3N}p}{h^{3N}} \rho_\mathrm{gc}\ln{\rho_\mathrm{gc}} 
= \tilde\beta\left\langle H \right\rangle - \mu\left\langle N \right\rangle + \ln{\mathcal{Z}}\, .
\label{g-canonical-4}
\end{align}
In general, we obtain, 
\begin{align}
{\mathcal{S}_0}^n = \left\langle \left(- \ln{\rho_\mathrm{gc}}\right)^n\right\rangle - \sum_{l=2}^{n}\frac{n!}{l!(n-l)!}\left(\tilde\beta\right)^l 
\sigma_l(H-\mu N)\left(\ln{\mathcal{Z}}\right)^{n-l}\, .
\label{g-canonical-10}
\end{align}
Here $\sigma_l (H - \mu N) = \left\langle (H - \mu N)^l\right\rangle - \left\langle H - \mu N\right\rangle^l$. 
Because
\begin{align}
\left\langle\left(H - \mu N\right)^l \right\rangle = \frac{1}{\mathcal{Z}} \sum_N \int \frac{d^{3N}q~d^{3N}p}{h^{3N}} \e^{-\tilde\beta(H-\mu N)} \left(H - \mu N\right)^l\, ,
\label{g-canonical-11}
\end{align}
the following expression can be obtained, 
\begin{align}
\sigma_l (H-\mu N) = \left(-1\right)^l \left\{\frac{1}{\mathcal{Z}}\frac{\partial^l\mathcal{Z}}{\partial\tilde\beta^l} 
 - \left(\frac{1}{\mathcal{Z}}\frac{\partial \mathcal{Z}}{\partial\tilde\beta}\right)^l\right\}\, .
\label{g-canonical-12}
\end{align}
We now define the following entropy in the grand canonical ensemble, 
\begin{align}
\mathcal{S}_\textrm{gr-can} = \sum_{n=0}^{\infty} \frac{f_\mathrm{g}^{(n)}}{n!}
\left\{\left\langle \left(-\ln{\rho_\mathrm{gc}}\right)^n\right\rangle - \sum_{l=2}^n \frac{n!}{l!(n-l)!}\left(\tilde\beta\right)^l\sigma_l(H-\mu N)\left(\ln{\mathcal{Z}}\right)^{n-l}\right\}\, .
\label{g-canonical-13}
\end{align}
Then we also obtain the expressions corresponding to (\ref{canonical-17}). 

In the grand canonical description, chemical potential corresponds to the work necessary to add a particle to the system by maintaining the equilibrium of the system. 
For the system to maintain the equilibrium, the particle must have a certain energy that is comparable to the mean energy of all the other particles. 

\section{Summary and Discussion}\label{SecIX}

In this review paper, we have first discussed if the Hawking temperature~\cite{Hawking:1974rv, Hawking:1975vcx} in (\ref{dS6BB}) 
(in the case of the Schwarzschild spacetime, we use (\ref{SchwrzP})) 
and the ADM mass \cite{Arnowitt:1959ah} could actually provide the thermodynamical temperature and energy uniquely. 

We have considered these problems in Section~\ref{SecII}. 
The Hawking temperature is given by the thermal distribution of the Hawking radiation, which is generated only by the geometry of the object
but does not depend on the details of the gravity theory. In this sense, the Hawking temperature is a unique possibility of the thermal temperature. 
About the ADM mass, if we consider the fall of the dust shell as a ``thought experiment'', as described in Section~\ref{SecIII}, 
by using energy conservation and Birkhoff's theorem~\cite{Wald:1984rg}, the thermodynamical energy must be given by the ADM mass. 
Then the thermodynamical relation $dE= T d\mathcal{S}$ tells us that the entropy of the system should be the Bekenstein-Hawking entropy~ \cite{Bekenstein:1973ur, Hawking:1974rv}. 

After that, in Section~\ref{SecIV}, we have explicitly checked if the generalised entropies could yield both the Hawking temperature and the ADM mass correctly. 
In particular, we have considered the R\'enyi entropy (\ref{RS1}) ~\cite{Czinner:2015eyk, Tannukij:2020njz, Promsiri:2020jga, Samart:2020klx}, in Section~\ref{SecIVA}, 
and the Tsallis entropy~(\ref{TS1})~\cite{Tsallis:1987eu}, in Section~\ref{SecIVB}. 
We have further investigated generalised entropies, like the four- and five-parameter generalised entropies, in (\ref{intro-1}) 
and (\ref{intro-4}) \cite{Nojiri:2022aof, Nojiri:2022dkr, Odintsov:2022qnn}
in Section~\ref{subge}. 

Despite the uniqueness of the Bekenstein-Hawking entropy, we consider the possibility that the generalised entropies could become true thermodynamical entropies. 
One possibility, which we discussed in Section~\ref{SecV}, is given by hairy black holes because the energy density of the hair contributes non-trivially to the ADM mass. 
We have considered the case of the Reissner-Nordstr\"{o}m black hole with the hair of the electric field, in Section~\ref{SecVA}, and the case of 
Einstein's gravity coupled with two scalar fields, in Section~\ref{SecVB}. 
By using the case of two scalar fields, we could realise an arbitrarily given spherically symmetric spacetime, which can be time-dependent in general \cite{Nojiri:2020blr}. 
The ghosts in the original model~\cite{Nojiri:2020blr} can be eliminated via some constraints~\cite{Nojiri:2023dvf}.
After providing some examples, in Section~\ref{SecVB2}, in the framework of the model with the two scalar fields, we have proposed two mechanisms to produce 
the generalised entropies in BH thermodynamics. 
In one case, Section~\ref{SecVB3}, we have investigated the possibility that, as in Reissner-Nordstr\"{o}m black hole, the horizon radius is not given only by the ADM mass 
and, therefore, the entropy becomes a non-trivial function of the Bekenstein-Hawking entropy, as shown for the R\'enyi entropy in (\ref{RSex3}) 
and for arbitrary generalised entropies in (\ref{gnrlM}). 
We have also considered the case where the effective mass expressing the energy inside the horizon does not give the naive Hawking temperature, 
as in (\ref{TH6}) of Section~\ref{SecVB4}. 
We have shown how the R{\'e}nyi entropy~(\ref{RS1}), the Tsallis entropy (\ref{TS1}), the three-parameter generalised entropy $\mathcal{S}_3$ (\ref{intro-3}), 
and the six-parameter entropy $\mathcal{S}_6$ (\ref{intro-2}) are generated in Eqs.~(\ref{entropy8}), (\ref{entropy6}), (\ref{general5h2}), and (\ref{general1h2}), respectively. 
Therefore, the inconsistency of new entropy proposals, with a Hawking temperature between the area law, could be avoided for the above black holes with one or more hair types. 
In \cite{Ma:2014qma}, the thermodynamical relations in the regular black holes were investigated and shown that the naive first law of thermodynamics 
using the Bekenstein-Hawking entropy is broken and the thermodynamical energy should be corrected by a factor. 
The reasons for the breakdown of the first law should be also due to the hair. 
The hairs coming from the electromagnetic fields, scalar fields, the Gauss-Bonnet terms, etc., outside the black hole horizon, 
contribute to the ADM mass as in the gravity theories coupled with the two scalar fields as investigated in this paper. 
Even in general modified gravity theories, there are hairs outside the horizon in general. 
These hairs contribute to the ADM mass and there occurs the breakdown of the naive first law. 

The radii of the photon sphere and of the black hole shadow have been calculated, for the models found in Section~\ref{SecVB}, in (\ref{ph2}) and (\ref{shphex}) 
and we obtained observational constraints on the parameters of the models in Section~\ref{generalshadow}. 
The parameters are consistent provided the BH is of the Schwarzschild kind. However, there is no direct relation between shadow and BH thermodynamics.

After that, we reviewed the generalised entropy description in the microcanonical, canonical, and grand canonical ensembles. 
The origins of the generalised entropies were discussed in Section~\ref{SecVII} in the formulations of a microcanonical ensemble, in Section~\ref{Sec2}, 
and of a canonical ensemble, in Section~\ref{Sec3}. 
After that, we used the McLaughlin expansion for the generalised entropies in Section~\ref{SecVIII} and possible interpretations were given. 

So far no observations exist to indicate the possibility that the BH entropy should be given by any of the non-extensive ones. 
But, eventually, future observations of black hole shadows, primordial gravitational waves from primordial black holes, as well as cosmological ones, 
might reveal significant discrepancies with Einsteinian gravity predictions. 
That would open the window for modified gravity theories and generalised entropies, which could correspond to the ones considered here. 
Until such observational results are obtained, it is important to be ready and to consider what kind of novel physical effects could appear thanks to the generalised entropies. 
In parallel, we need to consider how a generalised entropy may follow from a more fundamental, possibly quantum, theory of gravity, like superstring theory. 
Finally, generalised statistics/entropy may provide new connections between BH thermodynamics, cosmology and information theory, for instance, 
via the Landauer principle \cite{Trivedi:2024qys, Odintsov:2024ipb}.

\acknowledgments{We are grateful to V.~Faraoni and T.~Paul who collaborated with us on different studies of generalised entropies.
This work is supported by the program Unidad de Excelencia Maria de Maeztu CEX2020-001058-M, Spain (EE and SDO).
}


\begin{thebibliography}{99}

\bibitem{Hawking:1974rv}
S.~W.~Hawking,
Nature \textbf{248} (1974), 30-31
doi:10.1038/248030a0

\bibitem{Hawking:1975vcx}
S.~W.~Hawking,
Commun. Math. Phys. \textbf{43} (1975), 199-220
[erratum: Commun. Math. Phys. \textbf{46} (1976), 206]
doi:10.1007/BF02345020

\bibitem{Bekenstein:1973ur}
J.~D.~Bekenstein,
Phys. Rev. D \textbf{7} (1973), 2333-2346
doi:10.1103/PhysRevD.7.2333


\bibitem{Nojiri:2022aof}
S.~Nojiri, S.~D.~Odintsov and V.~Faraoni,
Phys. Rev. D \textbf{105} (2022) no.4, 044042
doi:10.1103/PhysRevD.105.044042
[arXiv:2201.02424 [gr-qc]].

\bibitem{Nojiri:2022dkr}
S.~Nojiri, S.~D.~Odintsov and T.~Paul,
Phys. Lett. B \textbf{831} (2022), 137189
doi:10.1016/j.physletb.2022.137189
[arXiv:2205.08876 [gr-qc]].

\bibitem{renyi}
A.~R{\'{e}}nyi, 
Proceedings of the Fourth Berkeley Symposium on Mathematics, Statistics 
and Probability, University of California Press (1960), 547-56

\bibitem{Tsallis:1987eu}
C.~Tsallis,
J. Statist. Phys. \textbf{52} (1988), 479-487
doi:10.1007/BF01016429


\bibitem{Ren:2020djc}
J.~Ren,
JHEP \textbf{05} (2021), 080
doi:10.1007/JHEP05(2021)080
[arXiv:2012.12892 [hep-th]].

\bibitem{Nojiri:2019skr}
S.~Nojiri, S.~D.~Odintsov and E.~N.~Saridakis,
Eur. Phys. J. C \textbf{79} (2019) no.3, 242
doi:10.1140/epjc/s10052-019-6740-5
[arXiv:1903.03098 [gr-qc]].


\bibitem{SayahianJahromi:2018irq}
A.~Sayahian Jahromi, S.~A.~Moosavi, H.~Moradpour, J.~P.~Morais Gra\c{c}a, I.~P.~Lobo, I.~G.~Salako and A.~Jawad,
Phys. Lett. B \textbf{780} (2018), 21-24
doi:10.1016/j.physletb.2018.02.052
[arXiv:1802.07722 [gr-qc]].

\bibitem{Barrow:2020tzx}
J.~D.~Barrow,
Phys. Lett. B \textbf{808} (2020), 135643
doi:10.1016/j.physletb.2020.135643
[arXiv:2004.09444 [gr-qc]].

\bibitem{Kaniadakis:2005zk}
G.~Kaniadakis,
Phys. Rev. E \textbf{72} (2005), 036108
doi:10.1103/PhysRevE.72.036108
[arXiv:cond-mat/0507311 [cond-mat]].

\bibitem{Drepanou:2021jiv}
N.~Drepanou, A.~Lymperis, E.~N.~Saridakis and K.~Yesmakhanova,
Eur. Phys. J. C \textbf{82} (2022) no.5, 449
doi:10.1140/epjc/s10052-022-10415-9
[arXiv:2109.09181 [gr-qc]].

\bibitem{Majhi:2017zao}
A.~Majhi,
Phys. Lett. B \textbf{775} (2017), 32-36
doi:10.1016/j.physletb.2017.10.043
[arXiv:1703.09355 [gr-qc]].


\bibitem{Arnowitt:1959ah}
R.~L.~Arnowitt, S.~Deser and C.~W.~Misner,
Phys. Rev. \textbf{116} (1959), 1322-1330
doi:10.1103/PhysRev.116.1322


\bibitem{Baruah:2024lyu}
A.~Baruah and P.~Phukon,
[arXiv:2411.02273 [hep-th]].

\bibitem{Shokri:2024elp}
M.~Shokri,
Phys. Lett. B \textbf{860} (2025), 139193
doi:10.1016/j.physletb.2024.139193
[arXiv:2411.00694 [hep-th]].

\bibitem{Sekhmani:2024frn}
Y.~Sekhmani, G.~G.~Luciano, S.~K.~Maurya, J.~Rayimbaev, B.~Pourhassan, M.~K.~Jasim and A.~Rincon,
Chin. J. Phys. \textbf{92} (2024), 894-914
doi:10.1016/j.cjph.2024.10.015

\bibitem{Lu:2024ppa}
H.~Lu, S.~Di Gennaro and Y.~C.~Ong,
[arXiv:2407.00484 [gr-qc]].

\bibitem{Xia:2024nmp}
J.~Xia and Y.~C.~Ong,
Universe \textbf{10} (2024) no.4, 177
doi:10.3390/universe10040177
[arXiv:2404.09278 [gr-qc]].

\bibitem{Barzi:2024bbj}
F.~Barzi, H.~El Moumni and K.~Masmar,
Nucl. Phys. B \textbf{1005} (2024), 116606
doi:10.1016/j.nuclphysb.2024.116606
[arXiv:2404.14609 [hep-th]].

\bibitem{Anand:2024txo}
A.~Anand and R.~Campos Delgado,
EPL \textbf{146} (2024) no.6, 69001
doi:10.1209/0295-5075/ad4c02
[arXiv:2403.13687 [gr-qc]].

\bibitem{Nakarachinda:2023jko}
R.~Nakarachinda, C.~Pongkitivanichkul, D.~Samart, L.~Tannukij and P.~Wongjun,
Fortsch. Phys. \textbf{72} (2024) no.7-8, 2400073
doi:10.1002/prop.202400073
[arXiv:2312.16901 [gr-qc]].

\bibitem{Gohar:2023hnb}
H.~Gohar and V.~Salzano,
Phys. Lett. B \textbf{855} (2024), 138781
doi:10.1016/j.physletb.2024.138781
[arXiv:2307.01768 [gr-qc]].

\bibitem{Cimidiker:2023kle}
I.~Cimidiker, M.~P.~D\c{a}browski and H.~Gohar,
Class. Quant. Grav. \textbf{40} (2023) no.14, 145001
doi:10.1088/1361-6382/acdb40
[arXiv:2301.00609 [gr-qc]].

\bibitem{Komatsu:2022bik}
N.~Komatsu,
Eur. Phys. J. C \textbf{83} (2023) no.8, 690
doi:10.1140/epjc/s10052-023-11855-7
[arXiv:2212.05822 [gr-qc]].

\bibitem{Nakarachinda:2022gsb}
R.~Nakarachinda, C.~Promsiri, L.~Tannukij and P.~Wongjun,
[arXiv:2211.05989 [gr-qc]].

\bibitem{Manoharan:2022qll}
M.~T.~Manoharan, N.~Shaji and T.~K.~Mathew,
Eur. Phys. J. C \textbf{83} (2023) no.1, 19
doi:10.1140/epjc/s10052-023-11202-w
[arXiv:2208.08736 [gr-qc]].

\bibitem{Cimdiker:2022ics}
I.~\c{C}imdiker, M.~P.~D\c{a}browski and H.~Gohar,
Eur. Phys. J. C \textbf{83} (2023) no.2, 169
doi:10.1140/epjc/s10052-023-11317-0
[arXiv:2208.04473 [gr-qc]].

\bibitem{DiGennaro:2022grw}
S.~Di Gennaro, H.~Xu and Y.~C.~Ong,
Eur. Phys. J. C \textbf{82} (2022) no.11, 1066
doi:10.1140/epjc/s10052-022-11040-2
[arXiv:2207.09271 [gr-qc]].

\bibitem{Afshar:2025sav}
M.~A.~S.~Afshar, M.~R.~Alipour, S.~Noori Gashti and J.~Sadeghi,
[arXiv:2501.00955 [gr-qc]].

\bibitem{NooriGashti:2024ywc}
S.~Noori Gashti, B.~Pourhassan and I.~Sakalli,
[arXiv:2412.12137 [hep-th]].

\bibitem{Nojiri:2021iko}
S.~Nojiri, S.~D.~Odintsov and T.~Paul,
Symmetry \textbf{13} (2021) no.6, 928
doi:10.3390/sym13060928
[arXiv:2105.08438 [gr-qc]].

\bibitem{Nojiri:2021jxf}
S.~Nojiri, S.~D.~Odintsov and T.~Paul,
Phys. Lett. B \textbf{825} (2022), 136844
doi:10.1016/j.physletb.2021.136844
[arXiv:2112.10159 [gr-qc]].


\bibitem{Li:2004rb}
M.~Li,
Phys. Lett. B \textbf{603} (2004), 1
doi:10.1016/j.physletb.2004.10.014
[arXiv:hep-th/0403127 [hep-th]].

\bibitem{Li:2011sd}
M.~Li, X.~D.~Li, S.~Wang and Y.~Wang,
Commun. Theor. Phys. \textbf{56} (2011), 525-604
doi:10.1088/0253-6102/56/3/24
[arXiv:1103.5870 [astro-ph.CO]].

\bibitem{Nojiri:2005pu}
S.~Nojiri and S.~D.~Odintsov,
Gen. Rel. Grav. \textbf{38} (2006), 1285-1304
doi:10.1007/s10714-006-0301-6
[arXiv:hep-th/0506212 [hep-th]].

\bibitem{Gong:2004cb}
Y.~g.~Gong, B.~Wang and Y.~Z.~Zhang,
Phys. Rev. D \textbf{72} (2005), 043510
doi:10.1103/PhysRevD.72.043510
[arXiv:hep-th/0412218 [hep-th]].

\bibitem{Khurshudyan:2016gmb}
M.~Khurshudyan,
Astrophys. Space Sci. \textbf{361} (2016) no.12, 392
doi:10.1007/s10509-016-2981-z

\bibitem{Landim:2015hqa}
R.~C.~G.~Landim,
Int. J. Mod. Phys. D \textbf{25} (2016) no.04, 1650050
doi:10.1142/S0218271816500504
[arXiv:1508.07248 [hep-th]].

\bibitem{Gao:2007ep}
C.~Gao, F.~Wu, X.~Chen and Y.~G.~Shen,
Phys. Rev. D \textbf{79} (2009), 043511
doi:10.1103/PhysRevD.79.043511
[arXiv:0712.1394 [astro-ph]].

\bibitem{Li:2008zq}
M.~Li, C.~Lin and Y.~Wang,
JCAP \textbf{05} (2008), 023
doi:10.1088/1475-7516/2008/05/023
[arXiv:0801.1407 [astro-ph]].


\bibitem{Nojiri:2019kkp}
S.~Nojiri, S.~D.~Odintsov and E.~N.~Saridakis,
Phys. Lett. B \textbf{797} (2019), 134829
doi:10.1016/j.physletb.2019.134829
[arXiv:1904.01345 [gr-qc]].


\bibitem{Bekenstein:1980jp}
J.~D.~Bekenstein,
Phys. Rev. D \textbf{23} (1981), 287
doi:10.1103/PhysRevD.23.287

\bibitem{Nojiri:2020blr}
S.~Nojiri, S.~D.~Odintsov and V.~Faraoni,
Phys. Rev. D \textbf{103} (2021) no.4, 044055
doi:10.1103/PhysRevD.103.044055
[arXiv:2010.11790 [gr-qc]].

\bibitem{Nojiri:2021czz}
S.~Nojiri, S.~D.~Odintsov and V.~Faraoni,
Phys. Rev. D \textbf{104} (2021) no.8, 084030
doi:10.1103/PhysRevD.104.084030
[arXiv:2109.05315 [gr-qc]].

\bibitem{Szabados:2009eka}
L.~B.~Szabados,
Living Rev. Rel. \textbf{12} (2009), 4
doi:10.12942/lrr-2009-4

\bibitem{Wald:1984rg}
R.~M.~Wald,
Chicago Univ. Pr., 1984,
doi:10.7208/chicago/9780226870373.001.0001

\bibitem{Odintsov:2022qnn}
S.~D.~Odintsov and T.~Paul,
Phys. Dark Univ. \textbf{39} (2023), 101159
doi:10.1016/j.dark.2022.101159
[arXiv:2212.05531 [gr-qc]].


\bibitem{Czinner:2015eyk}
V.~G.~Czinner and H.~Iguchi,
Phys. Lett. B \textbf{752} (2016), 306-310
doi:10.1016/j.physletb.2015.11.061
[arXiv:1511.06963 [gr-qc]].

\bibitem{Tannukij:2020njz}
L.~Tannukij, P.~Wongjun, E.~Hirunsirisawat, T.~Deesuwan and C.~Promsiri,
Eur. Phys. J. Plus \textbf{135} (2020) no.6, 500
doi:10.1140/epjp/s13360-020-00517-2
[arXiv:2002.00377 [gr-qc]].

\bibitem{Promsiri:2020jga}
C.~Promsiri, E.~Hirunsirisawat and W.~Liewrian,
Phys. Rev. D \textbf{102} (2020) no.6, 064014
doi:10.1103/PhysRevD.102.064014
[arXiv:2003.12986 [hep-th]].

\bibitem{Samart:2020klx}
D.~Samart and P.~Channuie,
Nucl. Phys. B \textbf{989} (2023), 116140
doi:10.1016/j.nuclphysb.2023.116140
[arXiv:2012.14828 [hep-th]].


\bibitem{Volovik:2024wif}
G.~E.~Volovik,
[arXiv:2409.15362 [physics.gen-ph]].


\bibitem{Misner:1964je}
C.~W.~Misner and D.~H.~Sharp,
Phys. Rev. \textbf{136} (1964), B571-B576
doi:10.1103/PhysRev.136.B571

\bibitem{Hernandez:1966zia}
W.~C.~Hernandez and C.~W.~Misner,
Astrophys. J. \textbf{143} (1966), 452
doi:10.1086/148525

\bibitem{Brown:1992br}
J.~D.~Brown and J.~W.~York, Jr.,
Phys. Rev. D \textbf{47} (1993), 1407-1419
doi:10.1103/PhysRevD.47.1407
[arXiv:gr-qc/9209012 [gr-qc]].

\bibitem{Nojiri:2022sfd}
S.~Nojiri, S.~D.~Odintsov and V.~Faraoni,
Int. J. Geom. Meth. Mod. Phys. \textbf{19} (2022) no.13, 2250210
doi:10.1142/S0219887822502103
[arXiv:2207.07905 [gr-qc]].

\bibitem{Bardeen:1973gs}
J.~M.~Bardeen, B.~Carter and S.~W.~Hawking,
Commun. Math. Phys. \textbf{31} (1973), 161-170
doi:10.1007/BF01645742

\bibitem{Page:2000dk}
D.~N.~Page,
[arXiv:hep-th/0012020 [hep-th]].

\bibitem{Nojiri:2023dvf}
S.~Nojiri and G.~G.~L.~Nashed,
Phys. Rev. D \textbf{108} (2023) no.12, 124049
doi:10.1103/PhysRevD.108.124049
[arXiv:2309.12379 [hep-th]].




\bibitem{Nojiri:2024dde}
S.~Nojiri, S.~D.~Odintsov and V.~Folomeev,
Phys. Rev. D \textbf{109} (2024) no.10, 104007
doi:10.1103/PhysRevD.109.104007
[arXiv:2401.15868 [gr-qc]].

\bibitem{Nojiri:2017kex}
S.~Nojiri and S.~D.~Odintsov,
Phys. Rev. D \textbf{96} (2017) no.10, 104008
doi:10.1103/PhysRevD.96.104008
[arXiv:1708.05226 [hep-th]].

\bibitem{Bambi:2019tjh}
C.~Bambi, K.~Freese, S.~Vagnozzi and L.~Visinelli,
Phys. Rev. D \textbf{100} (2019) no.4, 044057
doi:10.1103/PhysRevD.100.044057
[arXiv:1904.12983 [gr-qc]].

\bibitem{Vagnozzi:2022moj}
S.~Vagnozzi, R.~Roy, Y.~D.~Tsai, L.~Visinelli, M.~Afrin, A.~Allahyari, P.~Bambhaniya, D.~Dey, S.~G.~Ghosh and P.~S.~Joshi, \textit{et al.}
Class. Quant. Grav. \textbf{40} (2023) no.16, 165007
doi:10.1088/1361-6382/acd97b
[arXiv:2205.07787 [gr-qc]].

\bibitem{Nojiri:2023ikl}
S.~Nojiri and S.~D.~Odintsov,
Phys. Lett. B \textbf{845} (2023), 138130
doi:10.1016/j.physletb.2023.138130
[arXiv:2304.09014 [gr-qc]].

\bibitem{Nauenberg:2002azf}
M.~Nauenberg,
Phys. Rev. E \textbf{67} (2003), 036114
doi:10.1103/PhysRevE.67.036114
[arXiv:cond-mat/0210561 [cond-mat.stat-mech]].

\bibitem{Yoneya:1989ai}
T.~Yoneya,
Mod. Phys. Lett. A \textbf{4} (1989), 1587
doi:10.1142/S0217732389001817

\bibitem{Nojiri:2023bom}
S.~Nojiri, S.~D.~Odintsov and T.~Paul,
Phys. Lett. B \textbf{847} (2023), 138321
doi:10.1016/j.physletb.2023.138321
[arXiv:2311.03848 [gr-qc]].


\bibitem{Ma:2014qma}
M.~S.~Ma and R.~Zhao,
Class. Quant. Grav. \textbf{31} (2014), 245014
doi:10.1088/0264-9381/31/24/245014
[arXiv:1411.0833 [gr-qc]].


\bibitem{Trivedi:2024qys}
O.~Trivedi,
[arXiv:2407.15231 [gr-qc]].

\bibitem{Odintsov:2024ipb}
S.~D.~Odintsov, T.~Paul and S.~SenGupta,
[arXiv:2409.05009 [gr-qc]].

\end{thebibliography}
\end{document}